\documentclass[12pt,epsf]{article}
\usepackage{graphicx,subfigure}  
\def\be{\begin{equation}}
\def\ee{\end{equation}} 
\def\bea{\begin{eqnarray}}
\def\eea{\end{eqnarray}} 
\def\ba{\begin{array}} 
\def\ea{\end{array}}

\def\de{\delta}

\def\la{\lambda} 
\def\La{\Lambda}

\def\ln{{\rm ln}}
 
\def\nin{\noindent} 
\def\nn{\nonumber} 

\begin{document}

\begin{center} {\Large{\bf Effective potential (in)stability and lower bounds on  
the scalar (Higgs) mass}}\\

\vspace*{0.8 cm}

Vincenzo Branchina\footnote{Vincenzo.Branchina@ires.in2p3.fr}\label{one}, 
Hugo Faivre\footnote{Hugo.Faivre@ires.in2p3.fr}\label{two}
\vspace*{0.4cm}

{\it IReS Theory Group, ULP and CNRS,} \\
\vspace*{0.02cm}
{\it 23 rue du Loess, 67037 Strasbourg, France}\\

\vspace*{1 cm}

{\LARGE Abstract}\\

\end{center}

\vspace*{0.5cm}

It is widely believed that the top loop corrections to the Higgs effective 
potential destabilise the electroweak (EW) vacuum and that, imposing stability, 
lower bounds on the Higgs mass can be derived. With the help of a scalar--Yukawa 
model, we show that this apparent instability  is due to the extrapolation of 
the potential into a region where it is no longer valid. Stability turns out to 
be an intrinsic property of the theory (rather than an additional constraint to 
be imposed on it). However, lower bounds for the Higgs mass can still be 
derived with the help of a criterium dictated by the properties of the potential
itself. If the scale of new physics lies in the $\rm{Tev}$ region, 
sizeable differences with the usual bounds are found. Finally, our results exclude 
the alternative meta-stability scenario, according to which we might be living 
in a sufficiently long lived meta-stable EW vacuum. 

\vspace*{0.5cm}

\section{Introduction}

The Standard Model (SM) of particle physics is a very successful theory 
which has received a great number of experimental confirmations. As is 
well known, however, it is not complete. Its scalar sector, in particular, 
poses deep (and so far unanswered) questions. 

The value of the Higgs mass is not fixed by the theory, it is a free 
parameter. 
Nevertheless, in order to get informations on this fundamental 
quantity, theorists have tried to exploit at best the properties of the 
scalar sector of the SM (or some of its extensions). 

Through the analysis of the scalar effective potential, upper and lower 
bounds on the Higgs mass, $m_{_H}$, have been obtained  as a function of the  
physical cutoff, the scale of new physics. The upper bounds come from 
the triviality of the quartic  coupling \cite{ham} (for an alternative 
point of view see \cite{maurizio}), the lower ones from the 
requirement that the EW vacuum be stable (or, at least, meta-stable)
\cite{cab, sher3, lind, sher1, lind2, jones, sher2, alta, quiro1, quiro2}. 

For the lower bounds, the analysis is performed with the 
help of the RG-improved effective potential, $V_{_{RGI}}(\phi)$. 
Due to the $t \overline t $ loop corrections, $V_{_{RGI}}$ 
bends down for $\phi$ larger than $v$, the EW minimum. Depending on 
the value of the physical parameters, the resulting potential can be 
either unbounded from below up to the Plank scale, or can rise up again 
after forming a new minimum which is 
typically deeper than the EW vacuum. The latter is then said to be 
meta-stable.
   
As the instability occurs for sufficiently large values of the field, 
$V_{_{RGI}}$ is approximated by keeping only the quartic term \cite{sher2}. 
Using standard notations: 

\be\label{vrgi}
V_{_{RGI}}(\phi) \sim \frac{\overline{\lambda}(\phi)}{24} \phi^4\, .
\ee

\nin
In Eq.(\ref{vrgi}), the dependence of $\overline\lambda(\phi)$  
on $\phi$ is essentially the same as that of the corresponding RG-improved 
quartic coupling constant, $\lambda(\mu)$, on the running scale 
$\mu$, so that the behaviour of the 
effective potential can be read out from the $\lambda(\mu)$ 
flow \footnote{As correctly pointed out in \cite{quiro2}, however, 
$\overline{\lambda}(\phi)$ contains also terms not contained in 
$\lambda(\mu)$. They 
are really negligible only for very large values of $\phi$.}.               

The bending of the potential is due to the quarks-Higgs Yukawa couplings, 
namely to the minus sign carried by the fermion loops. Practically, it is 
sufficient to consider only the top, as the other (much lighter) quarks 
give comparably negligible contributions. 

The physical request that the EW vacuum 
be stable against quantum fluctuation is seen as an {\it additional 
phenomenological constraint}~  to be imposed on the effective 
potential. This constraint induces a relation between the physical cutoff 
and the Higgs mass. 

The derivation of the lower bounds goes as follows. 
Taking a boundary value for $\lambda(\mu)$ and for the 
other couplings, typically  
at $\mu=M_{_Z}$, the coupled RG equation are runned. As $\mu$ increases,
$\lambda(\mu)$ (initially) decreases. Depending on its initial value, 
$\lambda(M_{_Z})$, it 
may happen that at a certain scale, $\mu=\Lambda$, 
the running coupling $\lambda$ vanishes, becoming negative for 
higher values of $\mu$. Requiring that the EW vacuum be stable, 
$\Lambda$ is interpreted as the physical cutoff of the theory, 
the scale where new physics appears. 
From the matching condition, which relates 
$m_{_H}$ to $\lambda(M_{_Z})$ (at the tree level it is  
$m^2_{_H}=\frac{\lambda(M_{_Z})}{3}\,v^2$), a lower bound for $m_{_H}$
as a function of $\Lambda$ is obtained. This is the stability bound.

The possibility of having a minimum deeper than the EW one is also 
considered. The argument is that, as far as the tunnelling time between 
the false (EW) and the true vacuum is sufficiently large compared 
to the age of the Universe, we may well be living in the meta-stable EW 
vacuum. In this case, meta-stability bounds on $m_{_H}$ are 
found \cite {sher3, sher1, isido}.

These results, however, are at odds with a property of the effective 
potential, $V_{eff}(\phi)$, which, as is well known, is a convex
function of its argument \cite{sim, curt, rivers}.
It is also known that, when the classical potential is not 
convex (the phenomenologically interesting case), at any finite order 
of the loop expansion, $V_{eff}$ does not enjoy of this 
fundamental property. 
Alternative non-perturbative methods of computing the effective  
potential, though, such as lattice simulations \cite{latti}, 
variational approaches \cite{noivar}, or suitable averages of the 
perturbative results \cite{wewu}, provide the proper convex shape. 
The Wilsonian RG approach also gives a non-perturbative convex 
approximation for $V_{eff}$ \cite{fuku, ring, tet1, alex, tet2}. 

One of the main goals of the present work is to show that $V_{eff}$
is {\it nowhere unstable}. Its apparent instability is due to 
an extrapolation to values of $\phi$ which lie 
beyond its region of validity. Naively, however, the instability seems to occur 
in a region of $\phi$ where perturbation theory can be trusted \cite{jones}
and this explains why previous analyses have missed this point \footnote{ 
In addition, the use of RG techniques, which enlarge 
the domain of validity of perturbation theory via the resummation of leading, 
next to leading, ... logarithms, leads to the believe that the 
derivation of this instability is theoretically sound \cite{jones}.}.  

We also show that, despite the convexity of the potential, actually
thanks to this property, lower bounds for the 
Higgs mass can still be derived. Nevertheless, they no longer come as a result 
of an additional phenomenological constraint on $V_{eff}$,
namely the requirement of stability, they are already encoded in the theory. 
As we shall see, if the scale of new 
physics lies in the $\rm{Tev}$ region, the difference between our bounds and 
those obtained with the help of the usual stability criterium becomes sizeable.
The meta-stability scenario, on the contrary, is definitely excluded.

Finally, in order to shed more light on this (often mistreated
and misunderstood) subject, we reconsider here  some popular 
arguments \cite{sher1, dann}, sometimes quoted as the resolution 
of the instability (convexity) problem, and show that they are 
(at least) misleading. In section 2 we mainly 
concentrate on this last point which gives a good introduction to 
the subject and provides further motivation for our analysis. 

To understand the origin of the instability, we do not need to consider 
the complete SM. The group and the gauge structure of the theory are 
not essential for its occurrence. As it is due to the top-Higgs coupling 
(actually to the minus sign carried by the $t\overline{t}$-loop), the 
same instability occurs in the simpler model of a scalar coupled to a 
fermion with Yukawa coupling. To illustrate our argument, it will be 
sufficient to limit ourselves to consider this model. The extension of 
our results to the SM is immediate. 

The instability of the scalar effective potential is the subject 
of many studies. The one-loop (or higher loops) and the RG-improved potential 
are computed with the help 
of dimensional regularization. We also begin by computing the effective 
potential of our model in the $\overline{\rm MS}$ scheme (section 3). 
However, as will become clear in the following, dimensional 
regularization cannot reveal (in fact it masks) the origin of 
the problem.

The flaw in the usual procedure will be uncovered with the help of more 
physical renormalization schemes, the momentum cutoff regularization and the 
Wilsonian RG method. Dimensional regularization is a very powerful scheme 
which directly gives the finite results of renormalised perturbation 
theory. These other schemes allow to better follow the 
steps for the derivation of the renormalised potential from the bare 
one. This will help in finding the origin of the instability problem.

While completing our paper, we noted that this issue was recently 
considered in \cite{kuti, holland}. Although our conclusions look 
similar to those reached by these authors, we believe that 
their work differs from our in some important aspects, worth to 
be discussed. A comparison will be presented in the conclusions.

The rest of the paper is organized as follows. In section 2 we show 
how the Bogolubov criterium of dynamical instability allows to reconcile
the convexity of $\Gamma_{eff}$ with the existence of a broken phase 
and how the broken phase Green's functions can be derived 
from (the convex) $\Gamma_{eff}$. Moreover, 
we show how the dynamical instability 
criterium can be implemented within the framework of the Wilsonian 
RG method. In section 3 we compute the $\overline{MS}$ one-loop and 
RG-improved effective potential for our model and see that they both 
are unstable. 
In section 4 the same problem is considered within the 
momentum cut-off regularization scheme. In section 5 we analyse the 
results of the previous section and show that the instability 
comes from an illegal extrapolation of the  
renormalised potential beyond its range of validity. In addition, 
consistently with the stability constraint, we consider a criterium
for finding the physical cutoff of the theory. In section 6 we apply 
this criterium to the SM, thus getting lower bounds on the Higgs mass 
as a function of the scale of new physics, and compare with previous 
results. In section 7 we reconsider the instability problem within 
the framework of the non-perturbative Wilsonian RG method. 
Section 8 is for the summary and for our conclusions.

\section{Broken phase and dynamical instability.}

Before starting the detailed study of our model, in the present 
section we carefully analyse   
some popular arguments \cite{sher1, dann}, often presented as the 
resolution of the instability problem, and show that they are misleading. 
Moreover, by combining the Bogolubov criterium of dynamical instability 
with the Wilsonian RG method, we shall
provide further support to our analysis.

In \cite{sher1, dann} the effective action, $\Gamma_{eff}[\phi]$, 
and the generating functional of the broken phase ${\rm {1PI}}$ vertex 
functions, $\Gamma_{\rm{_{1PI}}}[\phi]$, are presented as two different 
functionals. Actually, these authors consider the first order in the 
$\hbar$-expansion of $\Gamma_{\rm{_{1PI}}}$, $\Gamma_{\rm{_{1PI}}}^{1l}$, 
and note that it is not convex. It is then argued that, when studying 
the stability of the EW vacuum, the relevant quantity to consider 
is $V_{_{\rm {_{1PI}}}}$ (or, more generally, 
its RG-improved version, $V_{_{\rm{RGI}}}$) rather than the convex 
$V_{eff}$, and that,  
being $V_{_{\rm {_{1PI}}}}$ non--convex,   
there is no convexity (instability) problem \cite{sher1}
\footnote{Presenting  
$\Gamma_{\rm {_{1PI}}}[\phi]$ and $\Gamma_{eff}[\phi]$
as two different quantities is a first source 
of confusion. As we have already said, the  convexity property of the exact 
$\Gamma_{eff}$ cannot be recovered within the loop expansion.
$\Gamma_{\rm{_{1PI}}}^{1l}$, which is the quantity considered in
\cite{sher1, dann}, is a non--convex, 
$O(\hbar)$, approximation of $\Gamma_{eff}$. It 
correctly approximates $\Gamma_{eff}$ in the neighbourhood of 
the minima (with some warnings specified later). 
In the region where it is non-convex, however, it is a bad approximation 
of $\Gamma_{eff}$.}. 

The argument is the following.  
 $V_{eff}(\phi)$ comes from the minimisation of 
$\langle\psi|\hat H|\psi \rangle$, where $\hat H$ is the 
energy density of the system and $|\psi \rangle$ is a state  which 
satisfies the constraint $\langle\psi|\hat\phi|\psi \rangle = \phi$.
For a symmetry breaking classical potential, the states 
that correspond to values of $\phi$ in the region between the classical 
minima, are not localised (more on this point later). As only localised states are 
of interest to us, and $V_{\rm{_{1PI}}}^{1l}$ is supposed to correspond 
to localised states also in the region between the minima \cite {wewu}, 
the conclusion is that $V_{_{\rm{1PI}}}^{1l}$ rather 
than $V_{eff}$ is the appropriate potential to consider. 

It is not difficult to see, however, that these lines of reasoning are 
misguiding. Indeed, the instability occurs for values of $\phi$ above $v$. 
Now, differently from those related to the region $-v \leq \phi \leq v$, 
the states that correspond to this range of $\phi$ are perfectly well 
localised and the above argument does not apply.

Moreover, as we shall briefly show below, the broken phase
zero momentum Green's functions, $\Gamma_n^{(v)}$, can be 
obtained from the convex $V_{eff}$ once we consider a physical 
procedure \cite{wit,stroc} based on the 
dynamical instability of the classical vacua (Bogolubov criterium) and  
that the usual loop expansion for 
$V_{eff}$ can be obtained within this framework.

This will help to further clarify the relation between $V_{eff}$ 
and $V_{_{\rm{1PI}}}^{1l}$. In any case, the potential to consider is 
$V_{eff}$, which is 
everywhere convex. However, as long as we are only interested in the broken 
theory Green's functions, i.e. in the local properties of $V_{eff}$ at 
$\phi=v$, it is possible (and from a practical point of view even more 
convenient) to consider a non-convex approximation, as 
$V_{_{\rm{1PI}}}^{1l}$ (or higher order ones), which coincides with 
$V_{eff}$ in the neighbourhood of $v$ (see below and footnote 5).

\begin{figure}[ht]
\begin{center}
\includegraphics[width=7cm,height=11cm,angle=270]{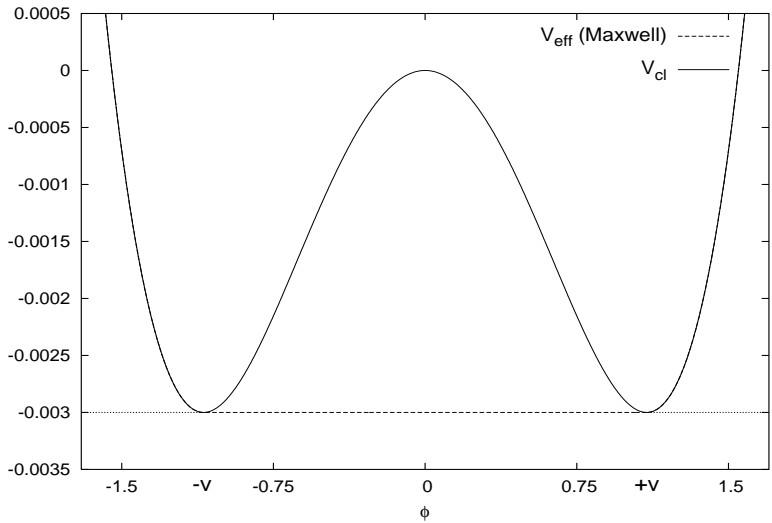}
\end{center}
\caption[]{The Maxwell construction for the classical potential
of the single component scalar theory considered in the text.
The parameters are chosen as:
$\lambda=5\cdot 10^{-2}$ and $m^2=-10^{-2}$.}
\label{fig1}
\end{figure}

Actually, the only range of $\phi$'s where a significative 
difference between the loop approximation and the exact effective 
potential is expected is the internal region, $-v \leq \phi \leq v$. 
The reason is easy to understand. By  construction, the one-loop 
approximation for the path integral which defines the effective potential 
considers the expansion of the action around a single saddle point. 
For values of $\phi$ in the internal region, however, there are two competing 
saddle points having the same weight \cite{rivers}. Taking into 
account both of these contributions, we get for the effective potential  
the known flat (convex) shape between the classical minima (Maxwell 
construction). On the contrary, for $\phi\geq v$ the path integral is 
dominated by a single saddle point. Therefore, no significative difference 
can occur in this region between the one-loop (or higher loop) approximation 
and the exact effective potential. 

A similar argument can be given within the framework of the Wilsonian 
RG approach where it was shown that, differently from 
the unbroken phase, the path integral which defines the infinitesimal 
RG-transformation for the Wilsonian potential in the broken phase 
is saturated by non-trivial saddle points \cite{alex}. 

Now we briefly show how the $\Gamma_n^{(v)}$'s are obtained from the convex 
effective action $\Gamma_{eff}$. For illustrational purposes, it is sufficient 
to consider the case of a constant background field, i.e. to consider $V_{eff}$ 
rather than the full effective action. Anyway, in the following, we are only 
interested in $V_{eff}$. For the sake of simplicity, we also limit ourselves  
to the case of a single component scalar theory.

\begin{figure}[ht]
\begin{center}
\includegraphics[width=7cm,height=11cm,angle=270]{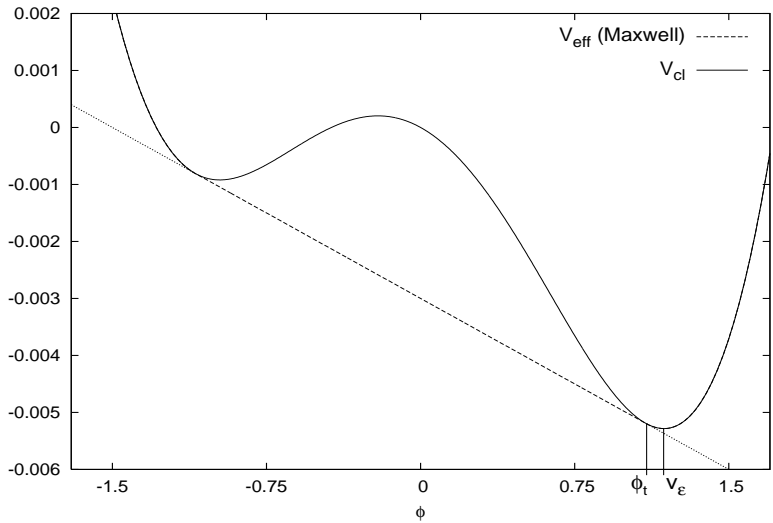}
\end{center}
\caption[]{The Maxwell construction for the classical potential
of the single component scalar theory with an explicit symmetry 
breaking term, $-\varepsilon \phi$. The parameters are chosen as:
$\lambda=5\cdot 10^{-2}$, $m^2=-10^{-2}$ and $\varepsilon=2\cdot 10^{-3}$.}
\label{fig2}
\end{figure} 

General theorems \cite{sim, ilyop}, together with several analytical 
and numerical non-perturbative 
studies \cite{latti, noivar, wewu, fuku, ring, tet1, alex, tet2}, 
indicate that  
$V_{eff}$ is a convex function of $\phi$ with a flat bottom 
between $-v$ and $v$, the minima of the classical potential. 
At the lowest order, $V_{eff}$ coincides with the well known
Maxwell (or double tangent) construction sketched in fig.\ref{fig1}.

The (zero momentum) $\Gamma_n^{(v)}$'s should be obtained by taking 
the derivatives of $V_{eff}$ at $\phi=v$. 
Due to the shape of the potential, however, this operation is
ambiguous and has to be defined with a certain care. 

The approach that we are going to consider now \cite{wit,stroc}, far from being 
a technical point, 
has a deep physical meaning. Following Bogolubov, in fact, we interpret the 
occurrence of symmetry breaking as a manifestation of the  
``dynamical instability" of the otherwise equivalent vacua 
of the potential. 
Adding to the Lagrangian an infinitesimal source term which explicitly 
breaks the classical symmetry of the theory, $-\varepsilon\phi$,  
we select one of the two classical vacua (see fig.\ref{fig2}). 
More precisely, this additional term creates an absolute 
minimum, $v_{_\varepsilon}$, close to the old $v$.  

As for the symmetric case, the lowest order for $V_{eff}$  
can be obtained with the help of the double tangent construction (fig.\ref{fig2}). 
A simple inspection of fig.\ref{fig2} shows that 
the derivatives at $\phi=v_{_\varepsilon}$ of the resulting 
modified effective potential, $V_{eff}(\phi \,; \varepsilon)$, can 
be safely taken. In fact, while in the symmetric case 
(fig.\ref{fig1}) the flat region extends from one of 
the classical minima to the other (the minima coincide with the
tangent points), in fig.\ref{fig2} the effective potential 
(as the classical one) has an absolute minimum, $v_{_\varepsilon}$, and  
the flat region starts at $\phi_{_t} < v_{_\varepsilon}$. 
The corresponding $\Gamma_n^{(v_{_\varepsilon}\, ;\, \varepsilon)}$'s
at this order are then easily obtained. The successive 
$\varepsilon\to 0$ limit \footnote{Although in this brief  
presentation we do not aim at complete rigour, it is worth to point out 
that to construct the $\Gamma_n^{(v)}$'s we begin first with a 
finite volume system and successively take the infinite volume limit. The 
latter has to be taken previous to the $\varepsilon\to 0$ limit.} 
gives the desired $\Gamma_n^{(v)}$'s.

Clearly, the $\Gamma_n^{(v)}$'s that we get this way 
are nothing but the usual tree level $\Gamma_n^{(v)}$'s. To get higher 
order approximations, we need to go beyond this lowest order Maxwell 
construction. Following \cite{alex}, we now show that, with the help 
of the Wilsonian RG approach, the above results can be established 
beyond this order.

As is well known, the non-perturbative RG equation for the Wilsonian 
effective potential, $U_{k}(\phi)$, in $d=4$ dimensions can be written 
as \cite{weg, nicol, hasen}:

\be\label{wils}
k\frac{\partial}{\partial k}U_{k}(\phi)= -\frac{k^4}{16\pi^2}
{\rm ln} \left(\frac{k^2 + U_{k}^{''}(\phi)}{k^2 + U_{k}^{''}(0)}\right)\, ,
\ee 

\nin
where the prime indicates derivation w.r.t. $\phi$. Note that the classical 
(bare) potential is $V_{cl}(\phi)= U_{_\Lambda}(\phi)$, while the effective potential 
is $V_{eff}(\phi)=U_{k=0}(\phi)$.

For a theory in the broken phase, however, Eq.(\ref{wils}) 
becomes unstable. More precisely, for values of $\phi$ in 
the internal region, this equation develops a singularity at 
finite critical values, $k_{cr}(\phi)$, of the running scale $k$.   
Starting from $k=k_{cr}(\phi)$, Eq.(\ref{wils}) is no longer valid. 

In \cite{alex} a new non-perturbative RG equation for $\phi$
in the unstable region was established:

\be\label{newrgeq}
U_{k-\delta k}(\phi)=min_{_{\{\varrho\}}} \left[ k^2 \varrho^2+ \frac{1}{2}
\int_{-1}^1\,dx\, U_k \Big(\phi+2 \varrho\cos(\pi x)\Big)\right] \, .
\ee

\nin
The minimum of Eq.(\ref{newrgeq}), $\varrho_{_k}(\phi)$, is the amplitude 
of the non-trivial saddle point which dominates the path integral  
defining the infinitesimal
RG-transformation ($k \to k -\delta k$) in the internal region. 
In the external region, on the contrary, the path integral is dominated 
by the trivial saddle point, i.e. $\varrho_{_k}(\phi)$ vanishes.

In \cite{alex}
the case of the symmetric potential (fig.\ref{fig1}) was considered and  
the Maxwell construction for $V_{eff}$ was established. Here we extend 
this analysis to the case of the potential with an explicit symmetry 
breaking term. 

\begin{figure}[ht]
\begin{center}
\includegraphics[width=7cm,height=11cm,angle=270]{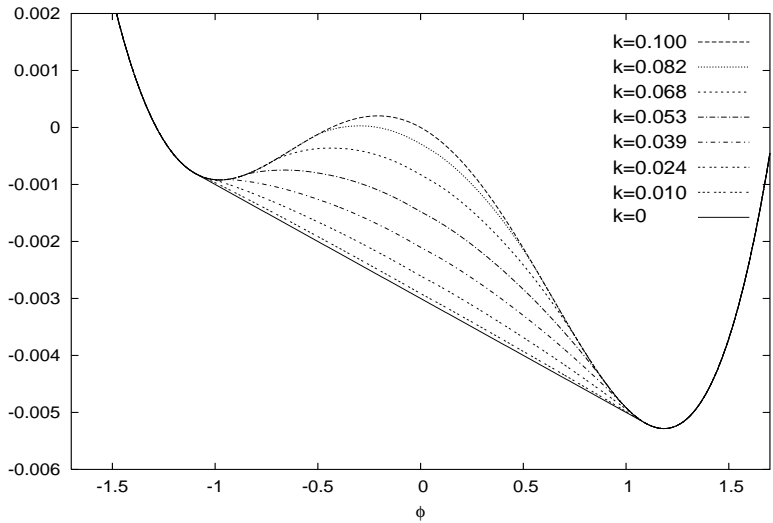}
\end{center}
\caption[]{RG flow for the potential of the single component scalar theory 
with explicit symmetry breaking term. Only the flow in the internal region 
is considered, i.e. the flow given by Eq.(\ref{newrgeq}). The boundary values 
for the parameters at $k=0.1$ are: $\lambda=5\cdot 10^{-2}$, $m^2=-10^{-2}$ and 
$\varepsilon=2\cdot 10^{-3}$ .}
\label{fig3}
\end{figure}

In fig.\ref{fig3} 
we show the flow of the Wilsonian potential, $U_k^{(\varepsilon)}(\phi)$, 
starting from the critical values $k_{cr}(\phi)$. From this 
figure we see that, even in the asymmetric case, there is a region where 
the effective potential, $V_{_{eff}}^{(\varepsilon)}(\phi) 
=U_{k=0}^{(\varepsilon)}(\phi)$, is flat and  
coincides with the double tangent construction. The same considerations 
done for the lowest order result are valid. In particular, 
the tangent point is displaced to the left of $v_{_\varepsilon}$ and 
the derivatives of $V_{_{eff}}^{(\varepsilon)}(\phi)$ 
at $v_{_\varepsilon}$ can be safely taken. 

The general conclusion of this analysis is that, with the help 
of Eqs.(\ref{wils}) and (\ref{newrgeq}), the 
Wilsonian potential can be runned all the way down from $k=\Lambda$ 
to $k=0$. The result is a non-perturbative convex approximation for $V_{eff}$ 
which shows the typical flat shape in the internal region (given by the
running of Eq.(\ref{newrgeq})), while in the external region has the shape 
governed by Eq.(\ref{wils}).   
 
We consider now the one-loop potential, 
$V^{1l}(\phi\, ; \varepsilon)$. In view of the previous discussion, 
it is not difficult to understand that, as far as we limit 
ourselves to consider a range of values of $\phi$ sufficiently 
close to the absolute minimum, $V^{1l}(\phi\, ; \varepsilon)$ 
provides a good approximation for $V_{eff}(\phi\, ; \varepsilon)$. 
Clearly, this is true for higher order loops too. 

Before ending this section, we would like to expand, as anticipated, 
on the argument according to which, when studying the 
stability of the vacuum, the convex $V_{eff}$ is not the 
appropriate potential to consider \cite{sher1}.

Let us indicate with $|v\rangle$ and $|-v\rangle$
the vacua constructed around $\phi =v$ and $\phi =-v$ respectively. 
The flatness of $V_{eff}$ in the $-v <\phi < v$ region 
implies that all the linear combinations of states  
$\alpha |v\rangle + \beta |-v\rangle$ (with $|\alpha|^2 + |\beta|^2 =1$)
are equivalent vacua, they all have 
the same energy. Apart from the trivial ones ($|\alpha|=1$, $\beta =0$
and $\alpha=0$, $|\beta| =1$), with any of the other non trivial combinations 
we would obtain Green's functions which violate the cluster decomposition
property. 
Moreover, the expectation value of the field is not constant allover $V$, the 
quantisation volume. In fact, for the generic 
state $\alpha |v\rangle + \beta |-v\rangle$, the expectation value  
$\langle\phi\rangle$ is
given by $(|\alpha|^2 - |\beta|^2) v$, and $V$ contains a fraction 
$|\alpha|^2$ of $\langle\phi\rangle = v$ and a fraction $|\beta|^2$ of 
$\langle\phi\rangle = - v$. Clearly, these states are not localised.

The above considerations are viewed as an indication that the convex 
$V_{eff}$ is not the appropriate potential to deal with.  
Although correct, these observations have nothing to do with the 
instability problem. As we have just seen, the non 
localised states correspond to values of $\phi$ in the internal region. 
The instability problem, however, occurs in the external region, where 
the states are perfectly well localised. 
Moreover, with the help of the Bogolubov criterium, we have seen how  
the degeneracy in the internal region is lifted and (in the 
infinite volume limit) only one vacuum is selected.

\section{One-loop and RGI potential. $\bf{\overline{MS}}$ Scheme.}

We compute now the one-loop effective potential, 
$V^{1l}$, for our model in the $\overline{MS}$ scheme and 
the corresponding RG-improved potential, $V_{_{RGI}}$.

The model consists of a single scalar field plus a single fermion field 
with scalar quartic interaction and Yukawa coupling, i.e.:

\be\label{lagra}
{\cal L}(\phi,\psi,\overline\psi)=\int\,d^4\,x\left(
\frac12\partial_\mu\phi\partial_\mu\phi 
+i\overline\psi\gamma_\mu\partial_\mu\psi+\frac{m^2}{2}\phi^2
+\frac\lambda2\phi^4 + g\phi\overline\psi\psi\right) .
\ee

\nin
Straightforward application of the $\overline{MS}$ prescriptions gives:   

\bea\label{Vdimreg}
V^{1l}(\phi)&=&\frac{1}{2}m^2\phi^2+\frac{\lambda}{24}\phi^4
+\frac{1}{64\pi^2}\left(m^2+\frac{\lambda}{2}\phi^2\right)^2\left({\rm ln}
\left(\frac{m^2+\frac{\lambda}{2}\phi^2}{\mu^2}\right)-\frac{3}{2}\right)\nn\\
&&-\frac{g^4\phi^4}{16\pi^2}\left({\rm ln}
\frac{g^2\phi^2}{\mu^2}   -\frac{3}{2}\right)\, ,
\eea

\nin
where $m^2$, $\lambda$ and $g$ depend on the renormalization scale $\mu$:

\be\label{cond}
m^2=m^2(\mu), ~~~~~\lambda=\lambda(\mu), ~~~~~ g=g(\mu).
\ee

In the r.h.s. of Eq.(\ref{Vdimreg}), the fermionic contribution 
comes with a negative sign. Therefore, we can easily find values of
$\lambda$ and $g$ (with $g^4 > \lambda$), together with a range of values 
of $\phi $, which satisfy the perturbative conditions,

\be\label{condizio}
\lambda < 1, \,\,\,\,\,\,  g < 1 \,\,\,\,\,\, {\rm and}\,\,\,\,\,\,  
\frac{g^4}{16\pi^2}{\rm ln}\frac{g^2\phi^2}{\mu^2} < 1 ,
\ee

\begin{figure}[Ht]
\begin{center}
\includegraphics[width=7cm,height=11cm,angle=270]{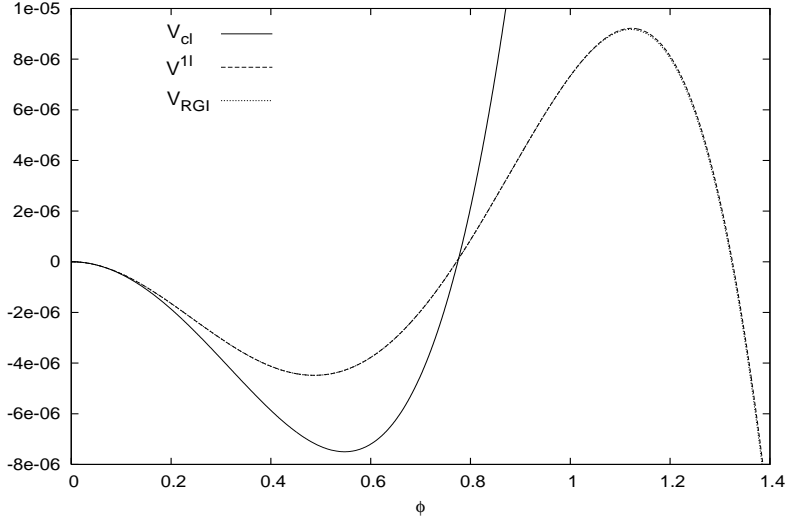}
\end{center}
\caption[]{Together with the classical potential, $V_{cl}$, of 
Eq.(\ref{lagra}), here we plot the one-loop, $V^{1l}$, and the 
RG-improved, $V_{_{RGI}}$, effective potential. The parameters 
are chosen at the scale
$\mu=1.1 \cdot 10^{-1}$ and are: $\lambda=2\cdot 10^{-3}$, 
$m^2=-10^{-4}$, $g=3\cdot 10^{-1}$. The instability of
$V^{1l}$ and $V_{_{RGI}}$ is immediately evident. Moreover, 
in this region, they are very close.}
\label{fig4}
\end{figure}

\nin
so that $V^{1l}(\phi)$ bends down and becomes 
lower than $V^{1l}(v)$ (see fig.\ref{fig4}).   
This is the instability problem for our one-loop potential. 

As is well known, we can improve on this result with the help
of renormalization group techniques. Let us consider the one-loop RG 
functions for $\lambda$, $g$, $m^2$ and for the vacuum 
energy \footnote{When considering the RG-improvement, 
the cosmological constant term has to be 
taken into account even if it was originally absent.} $\Omega$:

\bea\label{beta1}
\beta_{\lambda}&=&\frac{3\lambda^2}{16\pi^2}-\frac{3g^4}{\pi^2} 
\,~~~~~~~~~ ; \,~~~~~~~~~ 
\beta_{g}=\frac{g^3}{8\pi^2}~~~~~~~\nn\\
\beta_{_\Omega}&=&\frac{\lambda m^4}{32\pi^2} \,~~~~~~~~~~~~~~~~~~; \,~~~~~~~ 
\gamma_{_{m^2}}=\frac{\lambda}{16\pi^2}\, .
\eea

The largest logarithmic correction in the r.h.s. of Eq.(\ref{Vdimreg})
comes from the last term (the fermion). According with  
the RG--improvement logic, we now choose 
the running variable $t$ so that we get rid of this term in the improved potential: 
$t=\frac{1}{2}\ln\frac{g^2\phi^2}{\mu^2}-\frac{3}{4}$. As usual, the running functions  
$\overline{\lambda}(t)$, 
$\overline{g}(t)$, 
$\overline{m^2}(t)$, and
$\overline{\Omega}(t)$
are defined as the solutions of the differential equations: 

\bea\label{beta2}
\frac{d\,\overline{\lambda}}{d\,t}&=&\beta_{\lambda}(\overline\lambda,\overline
g,\overline\Omega,\overline m^2 ) \,~~~~~~~~ ; \,~~~~~~~
\frac{d\,\overline g}{d\,t}=\beta_{g}(\overline\lambda,\overline
g,\overline\Omega,\overline m^2 )~~~~\nn\\
\frac{d\,\overline{m}^2}{d\,t}&=&\gamma_{{m}^2} (\overline\lambda,\overline
g,\overline\Omega,\overline m^2 ) ~~~~~~~ ; ~~~~~~~
\frac{d\,\overline{\Omega}}{d\,t}=\beta_{\Omega}(\overline\lambda,\overline
g,\overline\Omega,\overline m^2 )
\eea 

\noindent
with boundary conditions: 

\be\label{beta3}
\overline{\lambda}(t=0)=\lambda \, ; ~~~\overline{g}(t=0)=g \,; ~~~
\overline{\Omega}(t=0)=0 \,; ~~~ \overline{m}^2(t=0)={m}^2 \, .
\ee

It is not difficult to see that the differential equations (\ref{beta2}) 
can be solved analytically. For $\overline g(t)$ and $\overline\lambda(t)$, 
for instance, we have:
\bea\label{analyt}
\overline g(t)&=&g\,\left(1-\frac{{g}^2\,t}{4\pi^2}\right)^{-\frac{1}{2}}\nn\\
\overline\lambda(t)&=&\frac{2}{3}\overline g^2(t)\left(1-\alpha+2\alpha 
{\left[1+{\left(\frac{\overline g(t)^2}
{{g}^2}\right)}^{\alpha}\,\frac{2{g}^2(\alpha+1)-3\lambda}
{2{g}^2(\alpha-1)+3\lambda}\right]}^{-1} \right)\, ,
\eea

\nin
with  $\alpha=\sqrt{37}$. 

Finally, the one-loop RG-improved potential is:

\be\label{improv}
V_{_{RGI}}=\frac{1}{2}\overline m^2(t)\phi^2+\frac{\overline \lambda(t)}{24}\phi^4
+\overline \Omega(t)+{\left( \frac{\bar m^2(t)+\frac{\bar{\la}(t)}{2}\phi^2}
{64\,\pi^2}\right)}^2\,\ln \frac{\bar m^2(t)+\frac{\bar{\la}(t)}{2}\phi^2}
{\overline g^2(t)\phi^2}.
\ee

In fig.~\ref{fig4} we plot $V_{_{RGI}}$ together with the one-loop and 
the classical potential for a particular choice of the renormalised parameters. 
A simple inspection of this figure shows that $V_{_{RGI}}$ (as well as 
$V^{1l}$) is unstable.   

\begin{figure}[Ht]
\begin{center}
\includegraphics[width=7cm,height=11cm,angle=270]{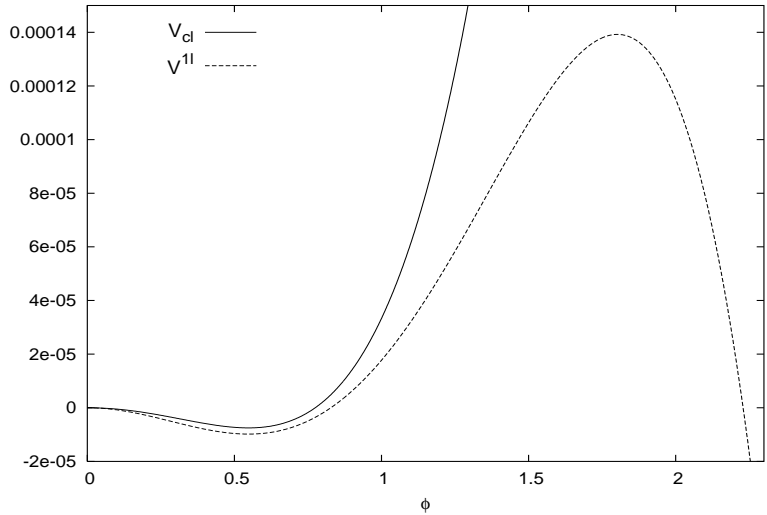}
\end{center}
\caption[]{Differently from Fig.4, here we have implemented the RG 
conditions so that the location of the minimum and the curvature of $V^{1l}$
at this point are the same as for $V_{cl}$ (see Appendix B). 
The parameters are chosen as in Fig.4.}
\label{fig5}
\end{figure}

Before ending this section, we would like to 
note that, due to the competition 
between the $\lambda^2$ and the $g^4$ terms in  $\beta_\lambda$ 
(first of Eqs.(\ref{beta1})), $\lambda (\mu)$, after decreasing for 
a certain range of energy, finally increases (toward the Landau pole). 
This generates a second 
minimum in the effective potential, typically lower than the first one.

Now, for certain values of $m_t$ and $m_{_H}$, 
which are compatible with the  current
experimental determinations and limits, the Higgs effective potential of 
the SM shows such a behaviour already below the Planck scale. 
As the tunnelling time between the false (EW) 
and the true vacuum appears to be sufficiently large (as compared to the age 
of the Universe), the alternative scenario of a meta-stable EW vacuum is 
also considered and lower 
meta-stability bounds on the Higgs mass are derived \cite{sher1,isido}. 
As we have anticipated, however, the proper treatment of the problem 
will show that effective potential is nowhere unstable. As a consequence, 
this scenario will be excluded.

\section{Momentum cutoff scheme}

In this section we show how the one-loop renormalised effective potential 
of Eq.(\ref{Vdimreg}) is obtained by considering the theory defined with 
a momentum cutoff. To prepare the discussion of the next section, we 
follow the computation in some detail. 

The parameters of the Lagrangian are now the bare ones. Therefore, 
in Eq.(\ref{lagra}) we replace $m^2$, $\lambda$ and $g$ with 
$m_{_{\Lambda}}^2$, $\lambda_{_{\Lambda}}$ and $g_{_{\Lambda}}$
respectively. As in the previous section, for the sake of simplicity, 
we neglect the wave function renormalization \footnote{When, in section 6, 
we shall be interested in the derivation of lower bounds on the Higgs 
mass, the anomalous dimension will be appropriately taken into account.}. 
A straightforward application of perturbation theory gives:

\bea\label{effec} 
V^{1l} (\phi)&=&\frac{m_{_{\Lambda}}^2}{2}\phi^2
+\frac{\lambda_{_{\Lambda}}}{24}\phi^4
+\frac{1}{64\pi^2}\Biggl\{\Lambda^4\,{\rm ln}
\left(\frac{\Lambda^2+m_{_{\Lambda}}^2+
\frac{\lambda_{_{\Lambda}}}{2}\phi^2}{\Lambda^2}
\right) ~~~~~~~~~~~~~~~~~~\nn\\
&+&\left(m_{_{\Lambda}}^2+\frac{\lambda_{_{\Lambda}}}{2}\phi^2\right)\Lambda^2  
-\left(m_{_{\Lambda}}^2+\frac{\lambda_{_{\Lambda}}}{2}\phi^2\right)^2 
{\rm ln}
\left(\frac{\Lambda^2+ m_{_{\Lambda}}^2+\frac{\lambda_{_{\Lambda}}}{2}\phi^2}
{ m_{_{\Lambda}}^2+\frac{\lambda_{_{\Lambda}}}{2}\phi^2}\right)\Biggr\}\nn\\ 
&-&\frac{1}{16\pi^2}\Biggl\{\Lambda^4\,{\rm ln}
\left(1+\frac{g_{_{\Lambda}}^2\phi^2}{\Lambda^2}
\right)+g_{_{\Lambda}}^2\phi^2\Lambda^2 -g_{_{\Lambda}}^4\phi^4{\rm ln}
\left(\frac{\Lambda^2+g_{_{\Lambda}}^2\phi^2}{g_{_{\Lambda}}^2\phi^2}\right)\Biggr\} . 
\eea

\nin
Considering only values of $\phi$ small compared to the cutoff, 

\be\label{condit} 
\frac{\phi}{\Lambda} < 1\, , 
\ee

\nin
expanding the r.h.s. of Eq.(\ref{effec}) in powers of 
$\frac{\phi}{\Lambda}$ and neglecting terms which are suppressed by 
negative powers of $\Lambda$, we get :

\bea\label{eff2} 
V^{1l} (\phi)~=~\frac{m_{_{\Lambda}}^2}{2}\phi^2
+\frac{\lambda_{_{\Lambda}}}{24}\phi^4
-\frac{1}{16\pi^2}\Biggl\{2g_{_{\Lambda}}^2\phi^2\Lambda^2 
-g_{_{\Lambda}}^4\phi^4\left[{\rm ln}
\left(\frac{\Lambda^2}{g_{_{\Lambda}}^2\phi^2}\right)+\frac12\right] \Biggr\}~~~ \nn \\
+\frac{1}{64\pi^2}\Biggl\{
2\left(m_{_{\Lambda}}^2+\frac{\lambda_{_{\Lambda}}}{2}\phi^2\right)\Lambda^2  
-\left(m_{_{\Lambda}}^2+\frac{\lambda_{_{\Lambda}}}{2}\phi^2\right)^2\left[ 
{\rm ln} \left(\frac{\Lambda^2}{m_{_{\Lambda}}^2+\frac{\lambda_{_{\Lambda}}}{2}\phi^2}
\right)+\frac12\right]\Biggr\} .
\eea 

We now move from bare to renormalised perturbation theory. 
After performing the splitting of the bare parameters in the usual way: 

\be\label{split}
m_{_{\Lambda}}^2=m^2+\delta m^2, ~~~~~\lambda_{_{\Lambda}}=\lambda+\delta \lambda, 
~~~~~ g_{_{\Lambda}}=g + \delta g, 
\ee

\nin
we insert Eq.(\ref{split}) in Eq.(\ref{eff2}) neglecting the higher order 
terms, i.e. removing $\delta m^2$, $\delta \la$ and $\delta g$ from  
the quantum fluctuation contribution. Finally, the counter-terms are determined 
so to cancel the quadratic and logarithmic divergences of $V^{1l}$. 

There is an arbitrariness in the determination of 
the counter-terms (different renormalization conditions) which is reflected 
in an arbitrariness in the finite parameters of 
the renormalised potential. By choosing: 

\bea\label{delta1}
\delta m^2&=& \delta m_{bos}^2 + \delta m_{fer}^2  \nn\\
\delta \lambda&=& \delta \lambda_{bos} + \delta \lambda_{fer} 
\eea

\nin
with ($\mu$ is an arbitrary low energy scale) 

\bea\label{delta2}
\delta m_{bos}^2 &=& -\frac{\lambda \Lambda^2}{32\pi^2}
                   +\frac{\lambda m^2}{32\pi^2}\left[
		   {\rm ln\left(\frac{\Lambda^2}{\mu^2}\right)} 
		   -1 \right] \,\, \, \,\, \,\,  ;\,\,\, \, \,\, \, \,
\delta m_{fer}^2 =\frac{g^2 \Lambda^2}{4\pi^2}~~~~~~~~~~\nn\\
\delta \lambda_{bos}&=&\frac{3\lambda^2}{32\pi^2}\left[
		   {\rm ln\left(\frac{\Lambda^2}{\mu^2}\right)} 
		   -1\right]
		   ~~~\, \, \, ;~\, \, \,
\delta \lambda_{fer}=-\frac{3 g^4}{2\pi^2}\left[
		   {\rm ln\left(\frac{\Lambda^2}{\mu^2}\right)}
		   -1\right] , 
\eea

\nin
we get:

\bea\label{Vcut}
V^{1l}(\phi)&=&\frac{1}{2}m^2\phi^2+\frac{\lambda}{24}\phi^4
+\frac{1}{64\pi^2}\left(m^2+\frac{\lambda}{2}\phi^2\right)^2\left({\rm ln}
\left(\frac{m^2+\frac{\lambda}{2}\phi^2}{\mu^2}\right)-\frac{3}{2}\right)\nn\\
&&-\frac{g^4\phi^4}{16\pi^2}\left({\rm ln}
\frac{g^2\phi^2}{\mu^2}   -\frac{3}{2}\right)\, ,
\eea

\nin
that is the one-loop potential of Eq.(\ref{Vdimreg}).

As for Eq.(\ref{Vdimreg}), the renormalised parameters that appear 
in Eq.(\ref{Vcut}) are defined at the scale $\mu$. Now, repeating the same 
steps of the previous section, we obtain from Eq.(\ref{Vcut}) the 
RG-improved potential of Eq.(\ref{improv}).

\section{Stability of the Effective Potential}

We show now that the effective potential is nowhere unstable,  
the claimed (apparent) instability being due to the 
extrapolation of $V^{1l}$ ($V_{_{RGI}}$) into a region where 
it is no longer valid. 

Before turning our attention to the renormalised potential, we begin 
by considering the bare theory as defined by the one-loop 
potential of Eq.(\ref{eff2}). For a certain region in the 
($m^2_{_\Lambda}$, $\lambda_{_\Lambda}$, $g_{_\Lambda}$)--parameter 
space, this potential, as the classical one, has two minima (Higgs phase). 
As we have already explained, the loop expansion is inadequate for the 
region between the minima. In the following, we ignore this region  
and concentrate our attention only on the external one, where the 
loop-expansion is expected to hold (as we know, in the internal 
region the convexity is restored via the Maxwell construction). 

\begin{figure}[Ht]
\begin{center}
\includegraphics[width=7cm,height=11cm,angle=270]{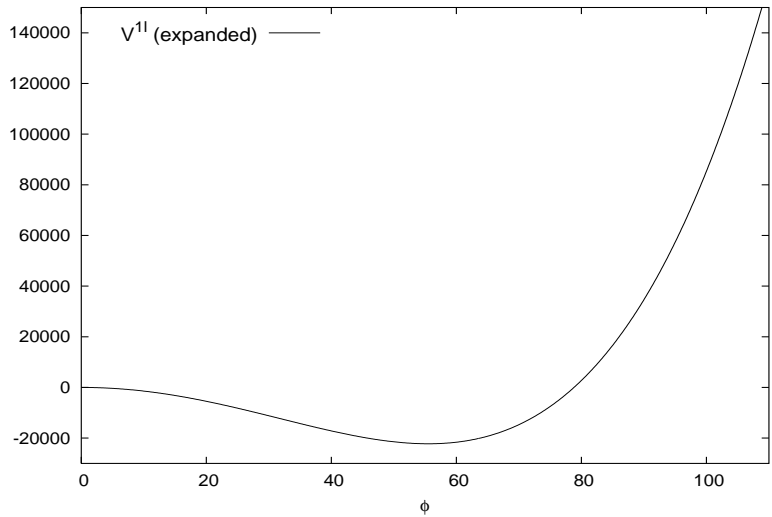}
\end{center}
\caption[]{The one-loop  effective potential of Eq.(\ref{eff2}) (before 
the subtraction of the quadratic divergences), for  
$\lambda_{_\Lambda}=5\cdot 10^{-2}$, $m^2_{_\Lambda}=-10^{-2}$, 
$g_{_\Lambda}=0.35$ and $\Lambda=100$. Neglecting, as explained in the text, 
the internal region, we see that beyond the minimum the potential is convex.}
\label{fig6}
\end{figure}

A careful analysis of Eq.(\ref{eff2}) shows that, in the 
external region, and within the range of $\phi$ where the potential is 
defined, i.e. for $\frac{|\phi|}{\Lambda} < 1$, the bare effective potential
is convex (in agreement with exact theorems). Therefore, it 
does not present any instability. 
In fig.\ref{fig6} we show a plot of $V^{1l}$, Eq.(\ref{eff2}),  
for a particular choice of the parameters. 

\begin{figure}[Ht]
\begin{center}
\subfigure[]{\includegraphics[width=7cm,height=6.5cm,angle=270]{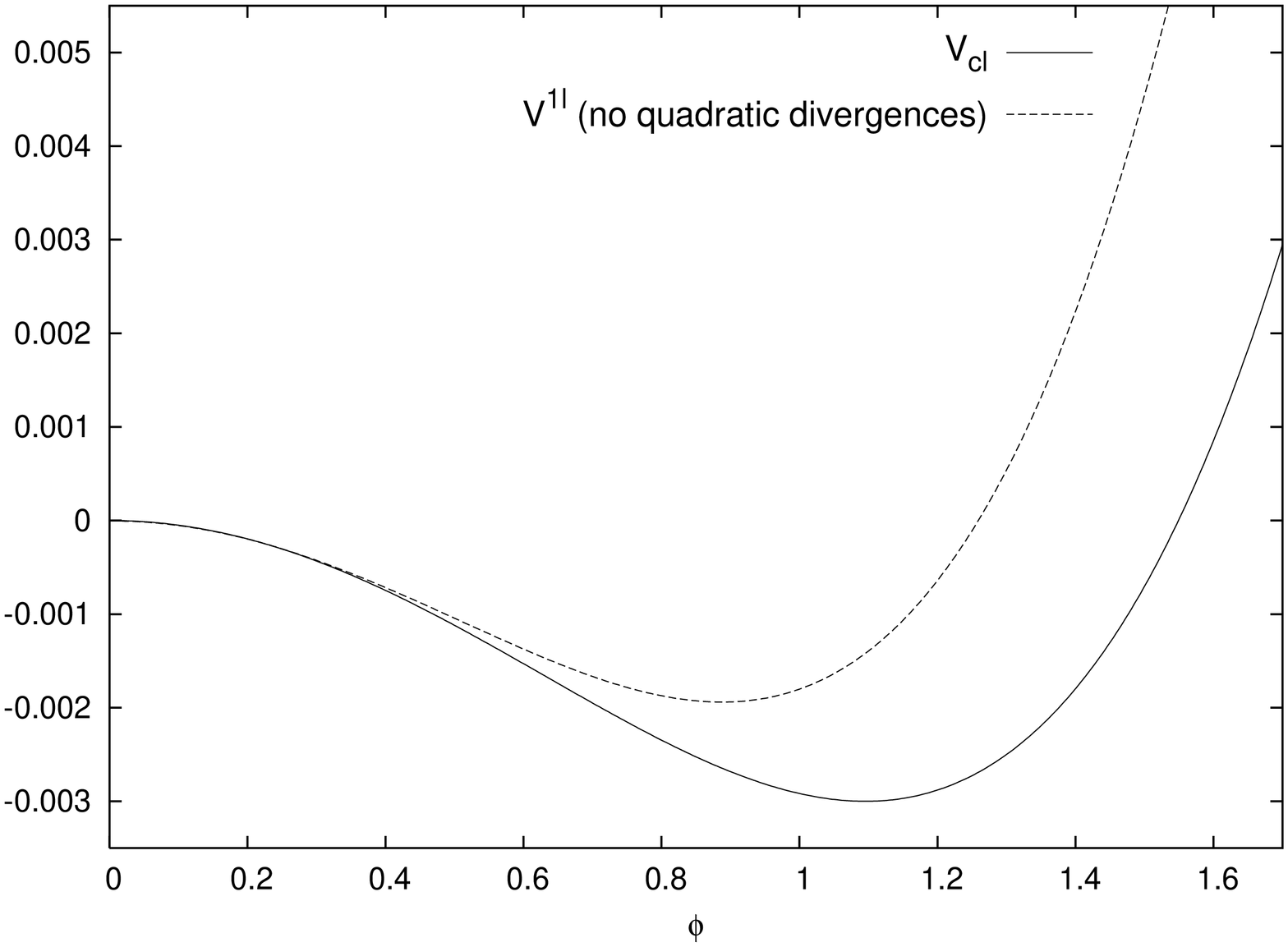}}
\subfigure[]{\includegraphics[width=7cm,height=6.5cm,angle=270]{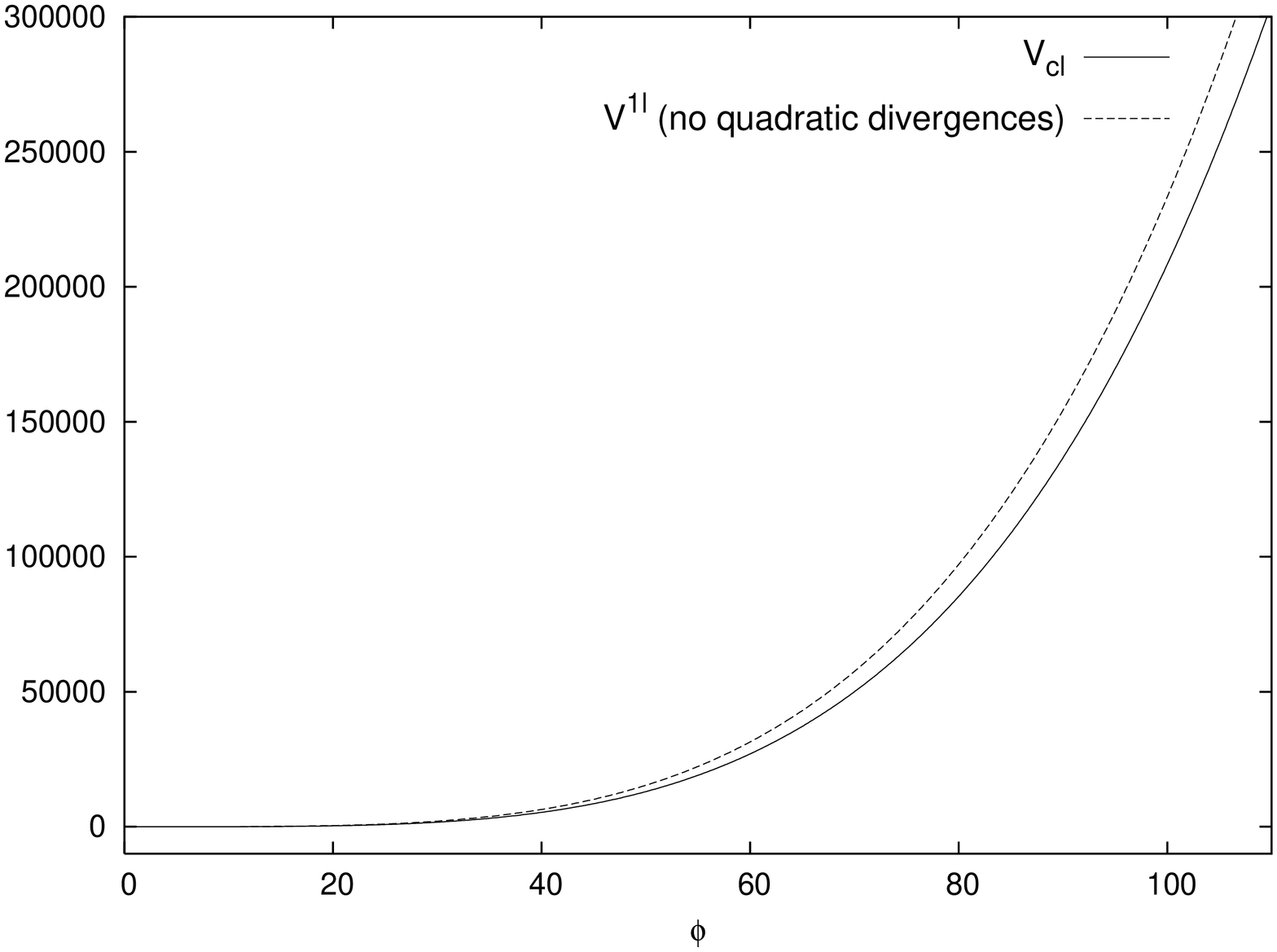}}
\end{center}
\caption[]{The bare potential together with the one-loop potential 
after subtraction of the quadratic divergences. The bare parameters are 
chosen as in Fig.5. (a) We zoom on a small
region of $\phi$, close to the classical minimum.  (b) Here we see that the 
effective potential, up to the cutoff scale, is stable. }
\label{fig7}
\end{figure}

We now subtract from the bare potential of 
Eq.(\ref{eff2}) the quadratically divergent terms \footnote{ 
As is well known, a well defined physical meaning can be attached to 
this operation. In a non-supersymmetric scenario, this 
cancellation is interpreted as the result of the conspiracy between 
unknown degrees of freedom, which live above the cutoff, and the quantum 
fluctuations of the fields below the cutoff. This way, the scalar (Higgs) 
mass is protected from getting too large corrections from the quantum 
fluctuations (this interpretation, however, poses the 
problem of the fine tuning required for the cancellation, the 
naturalness problem). In a susy scenario, on the contrary, 
this cancellation is obtained in a more ``natural" way. It is due to 
the presence of additional degrees of freedom (fields) below the cutoff.}. 
As illustrated in fig.\ref{fig7} for a specific
choice of the parameters, again the resulting potential turns out to be 
convex (once more, only the external region has to be considered). 
The bare potential, even after the subtraction of the quadratic 
divergences, does not show any sign of instability.

As an aside remark, we note that, as they describe different degrees of freedom,
the potentials of figs.\ref{fig6} and \ref{fig7} actually belong to two 
different effective theories (with or without the $\Lambda^2$ terms). 
From the point of view of the phenomenological applications in particle 
physics, however, we are typically interested in the theory where the 
quadratically divergent terms are subtracted. 

We have just seen that the bare potential (before and after the subtraction 
of the quadratic divergences) is everywhere stable. How can the renormalised 
potential show an instability? To answer this question, let us consider again 
Eq.(\ref{eff2}) after the subtraction of the quadratically 
divergent terms. 

As we already know, the instability occurs because 
the quantum fluctuations due to the fermions can compensate and then overhelm 
the classical $\phi^4$ term. Therefore, with no loss of generality, 
we can now neglect the bosonic contribution, as well as other unimportant 
finite terms, and limit ourselves to write:

\be\label{Vap2}
V^{1l}(\phi)=\frac12 m_{_{\Lambda}}^2\phi^2+
\frac{\lambda_{_{\Lambda}}}{24}\phi^4+\frac{g_{_{\Lambda}}^4\phi^4}{16\pi^2}{\rm ln}
\frac{\Lambda^2}{g_{_{\Lambda}}^2\phi^2} .
\ee

\nin
At a lower scale $\mu$ ($< \Lambda$) we have:

\be\label{Vap3}
V^{1l}(\phi)=\frac12 m_{_{\mu}}^2\phi^2+
\frac{\lambda_{_{\mu}}}{24}\phi^4+\frac{g_{_{\mu}}^4\phi^4}{16\pi^2}{\rm ln}
\frac{\mu^2}{g_{_{\mu}}^2\phi^2} \, ,
\ee

\nin 
which is the same potential of Eq.(\ref{Vap2}) written in terms of the renormalised 
parameters $m_{_{\mu}}^2$, $\lambda_{_{\mu}}$ and $g_{_{\mu}}$.

Clearly, if Eq.(\ref{Vap2}) does not show any instability, the same is true 
for Eq.(\ref{Vap3}). However, let us pretend (for the moment) that we have 
not made this observation and move to consider the usual phenomenological 
application of Eq.(\ref{Vap3}), which could have been obtained (section 3)
within the $\overline{MS}$ scheme.

From the first two terms of Eq.(\ref{Vap3}),  the classical vacuum, 

\be
v^2=-\frac{6m_{_{\mu}}^2 }{\lambda_{_{\mu}}} , 
\ee

\nin
is obtained. 
The last term can destabilise this vacuum if it becomes too large and 
negative. Strictly speaking, the presence of the last term also modifies
the position of the classical minimum, but this is not a  
complication. In fact, although this is not a necessary step, we can 
slightly modify the above 
expressions by adopting renormalization conditions that keep the 
position of the minimum unchanged. With this choice, 
Eq.(\ref{Vap3}) is replaced by (see Appendix A):

\be\label{newrg}
V^{1l}(\phi)=\frac12 m_{_v}^2\phi^2+
\frac{\lambda_{_v}}{24}\phi^4-\frac{g_{_v}^4\phi^4}{16\pi^2}
\left({\rm ln}\frac{\phi^2}{v^2} -\frac32 \right)
-\frac{g_{_v}^4 v^2}{8\pi^2}\phi^2.
\ee

In Eq.(\ref{newrg}) we have defined the parameters $m^2_{_v}$, 
$\lambda_{_v}$ and $g_{_v}$ at the IR scale $v$, the classical 
(and quantum) minimum, which is now given by: 

\be\label{newv}
v^2=-\frac{6~m_{_{v}}^2 }{\lambda_{_{v}}} . 
\ee

\nin
Correspondingly, Eq.(\ref{Vap2}) is replaced by:  

\be\label{newrg2}
V^{1l}(\phi)=\frac12 m_{_\Lambda}^2\phi^2+
\frac{\lambda_{_\Lambda}}{24}\phi^4-\frac{g_{_\Lambda}^4\phi^4}{16\pi^2}
\left({\rm ln}\frac{\phi^2}{\Lambda^2} -\frac32 \right)
-\frac{g_{_\Lambda}^4 v^2}{8\pi^2}\phi^2.
\ee

Going back to Eq.(\ref{newrg}), we now look for values of 
$\lambda_{_v}$, $g_{_v}$ and $\phi$ such that this equation 
is (expected to be) valid and, at the same time, give: 

\be\label{lower}
V^{1l}(\phi) < V^{1l}(v) \,.
\ee

\nin
The usual requirements for the validity  Eq.(\ref{newrg}) are that 
the renormalised coupling constants, $\lambda_{_v}$ and $g_{_v}$, 
as well as  the quantum correction, 
$\frac{g_{_v}^4}{16\pi^2}{\rm ln}\frac{\phi^2}{v^2}$, be perturbative,
i.e.: 

\be\label{pertu} 
\lambda_{_v} < 1 ~~~~~~~ , ~~~~~~~ g_{_v} < 1 \, , 
\ee

\nin
and 

\be\label{pertur} 
\left |\frac{g_{_v}^4}{16\pi^2}{\rm ln}\frac{\phi^2}{v^2}\right |< 1 
\ee

\nin
(note that Eqs.(\ref{pertu}) and (\ref{pertur}) are nothing but the perturbative
conditions of Eq.(\ref{condizio}) adapted to our current choices). 

In the following we show that, contrary to the common expectation, 
Eqs.(\ref{pertu}) and (\ref{pertur}) are not sufficient to garantee that 
Eq.(\ref{newrg}) can be trusted. An additional condition has to be considered. 
As we shall see, the apparent instability of the potential is due to the 
neglect of this condition. 

Let us choose $\lambda_{_v}$ and $g_{_v}$ so that these couplings, 
in addition to Eq.(\ref{pertu}), also satisfy the relation:

\be\label{rela}
\lambda_{_v}= \frac{3g_{_v}^4}{4\pi^2} . 
\ee 

\nin
Moreover, let us consider $\overline\phi$ such that:

\be\label{logaro}
{\rm ln}\frac{\overline\phi^2}{v^2} = 2 .
\ee 

Being $g_{_v} < 1$, it is a trivial exercise to see that, 
by virtue of Eq.(\ref{logaro}), Eq.(\ref{pertur}) holds for 
$\overline\phi$. Moreover, inserting 
Eqs.(\ref{rela}) and (\ref{logaro}) in Eq.(\ref{newrg}), we find: 

\be
V^{1l}(\overline\phi) < V^{1l}(v) . 
\ee

We would conclude that, in the range of $\phi$ given by 
$v<\phi<\overline\phi$, the renormalised potential  
of Eq.(\ref{newrg}) can be trusted and its instability 
(see fig.\ref{fig4}) is theoretically well established. 
In fact, this is what is usually 
stated \cite{jones}. In order to avoid any misunderstanding, 
it is worth to stress that the RG-improvement cannot change 
this conclusion. In the range of $\phi$ that we consider 
here, the condition (\ref{pertur}) holds so that, in this region, 
$V^{1l}$ and $V_{_{RGI}}$ are very close one to the other . 
  
As solid as they can seem, however, the above conclusions are 
incorrect. 

\nin
To understand why, let us first simplify (without any 
loss of generality) the discussion 
by neglecting in the following the running of $m^2$ and $g$. 
From Eqs.(\ref{newrg}) and (\ref{newrg2}) we have then:

\be\label{newrg3}
\frac{\lambda_{_{\Lambda}}}{24}\phi^4 + \frac{g^4\phi^4}{16\pi^2}\,{\rm ln}
\frac{\Lambda^2}{\phi^2} = 
\frac{\lambda_{_{v}}}{24}\phi^4 + \frac{g^4\phi^4}{16\pi^2}\,{\rm ln}
\frac{v^2}{\phi^2} ,
\ee

\nin
which immediately gives:

\be\label{rg}
\lambda_{_{\Lambda}} = \lambda_{_{v}} -\frac{3 g^4}{2\pi^2}{\rm ln}
\frac{\Lambda^2}{v^2} .
\ee

\nin
Inserting now Eqs.(\ref{rela}) and (\ref{logaro}) in Eq.(\ref{newrg3}), we find:

\be\label{impos}
\frac{\lambda_{_{\Lambda}}}{24} + \frac{g^4}{16\pi^2}\,{\rm ln}
\frac{\Lambda^2}{\overline\phi^2} =
\frac{\lambda_{_{v}}}{24} + \frac{g^4}{16\pi^2}\,{\rm ln}
\frac{v^2}{\overline\phi^2} < 0 .
\ee
 
\nin
Naturally, for the theory to be defined, it is $\lambda_{_{\Lambda}} > 0$.
Therefore, in order for Eq.(\ref{impos}) to 
be valid, we should have:

\be\label{contra}
\frac{\Lambda^2}{\overline\phi^2} \leq 1 .
\ee

Eq.(\ref{contra}) shows that, contrary to our naive expectation,  
$\overline\phi$ lies beyond the range of validity of $V^{1l}$  
($V_{_{RGI}}$).

We now understand the origin of the apparent instability of 
the renormalised potential. If, in order to decide whether a certain 
value of $\phi$ belongs to the region where $V^{1l}$ can be trusted, 
we only consider Eqs. (\ref{pertu}) and (\ref{pertur}), {\it we loose 
the information contained in the additional independent condition}: 

\be\label{newrg4}
\frac{\lambda_{_{v}}}{24}\phi^4 + \frac{g^4\phi^4}{16\pi^2}\,{\rm ln}
\frac{v^2}{\phi^2} > 0 \, .
\ee

When, on the contrary, this condition is taken into account, 
the effective potential does not present any instability. In 
other words, the instability occurs in a region of $\phi$'s
where Eq. (\ref{newrg}) for $V^{1l}$ is no longer valid. 

Naturally, these same conclusions could have been reached by looking at 
the problem the other way around. In fact, coming back to the 
observation that we have put aside before, we note that, due to 
the condition $\phi^2 < \Lambda^2$, the 
combination $\frac{\lambda_{_{\Lambda}}}{24} + \frac{g^4}{16\pi^2}{\rm ln}
\frac{\Lambda^2}{\phi^2}$ cannot be negative. Therefore, Eq.(\ref{impos}) 
cannot be fulfilled and no instability can occur.

The above longer discussion, however, is motivated by
the common believe that, in order to ascertain the validity of
the result for $V^{1l}$, Eqs. (\ref{pertu}) and (\ref{pertur}) are 
the {\it only conditions} to be verified. Actually, this is 
the reason why it is still believed that the instability of $V^{1l}$ 
(and $V_{_{RGI}}$) is a genuine effect due to the quantum corrections.

We can now deepen our analysis by noting that, 
as an elementary exercise shows, 
the point beyond the minimum where the effective potential ceases to 
be convex, i.e. the inflection point in the external region, 
$\phi_{inf}$, is such that:

\be\label{inflect}
\phi_{inf} \geq \Lambda .
\ee

Eq.(\ref{inflect}) is important for two reasons. On the one hand, it  
shows that the effective potential is convex wherever it 
is defined. On the other hand, it provides a 
criterium for the derivation of lower bounds on the scalar (Higgs) mass. 

To better understand this last point, let us consider the usual approach, 
where a bound on the renormalised $\lambda$ is obtained from (the 
equivalent of) Eq.(\ref{rg}). 
At first it is noted that the instability occurs if  
$V^{1l}(\phi_0) = V^{1l}(v)$ at a certain $\phi_0$ and 
$V^{1l}(\phi) < V^{1l}(v)$ for $\phi > \phi_0$. Then it is shown that 
$\phi_0$ (almost) corresponds to the value of the running scale where 
$\lambda (\mu)$ vanishes (see Eq.(\ref{vrgi}) and footnote 3). 
Finally, a vanishing $\lambda_{_\Lambda}$
is taken in Eq.(\ref{rg}) so that the highest possible physical 
cutoff $\Lambda$, corresponding to a given value of the renormalised coupling 
$\lambda_v$, is derived.

Instead, our analysis suggests that the upper bound for the range 
of $\phi$'s, which is also the highest self-consistent value for the 
physical cutoff, should be taken at the inflection point of 
Eq.(\ref{inflect}), the value of $\phi$ where the potential ceases to 
be convex.

Although up to now we have considered a simple scalar-Yukawa model, it is clear
that our results are completely general. In the next section we shall see 
how the above criterium 
can be exported into the SM to get lower bounds on the Higgs mass.
 
Before we move to this phenomenological application, however, it is worth 
to stress that in the usual approach the requirement of stability appears 
to be an extra phenomenological constraint to be possibly imposed on 
the theory; an unstable potential is considered as a 
legitimate one. In fact, the meta-stability scenario, clearly excluded by 
our analysis, is based on the possibility of having a second minimum of 
the potential lower that the EW vacuum. As we have seen, however, 
the stability of the effective potential is an intrinsic property of the 
theory. No place is left for an unstable or meta--stable potential.

\section{Lower bounds on the Higgs mass}

Let us consider now some important phenomenological implications 
of our findings for the SM. 
Clearly, the first thing to point out is that, contrary to common 
believe, the Higgs effective potential does not present any instability.  
As for the determination of the lower bounds on $m_{_H}$, we have 
seen that the internal consistency of the theory 
requires that the physical cutoff has to be taken at the location 
of the inflection point of the potential (in the region 
beyond the minimum). 

Implementing this criterium for the determination of the scale of new 
physics, lower bounds for the Higgs mass are found. Our results 
will be compared with those obtained with the help of the 
usual instability criterium. 

The well known one-loop potential of the scalar sector 
of the SM reads \cite{cw}:

\bea
V^{1l}(\phi)=\frac{1}{2}m^2\phi^2+\frac{\lambda}{24}\phi^4
+\frac{1}{64\pi^2}\Bigg[ \left(m^2+\frac{\lambda}{2}\phi^2\right)^2\left({\rm ln}
\left(\frac{m^2+\frac{\lambda}{2}\phi^2}{\mu^2}\right)-\frac{3}{2}\right)\nn\\
+3\,\left(m^2+\frac{\lambda}{6}\phi^2\right)^2\left({\rm ln}
\left(\frac{m^2+\frac{\lambda}{6}\phi^2}{\mu^2}\right)-\frac{3}{2}\right)
+6\,\frac{{g_1}^4}{16}\phi^4\left({\rm ln}
\left(\frac{\frac{1}{4}{g_1}^2\phi^2}{\mu^2}\right)-\frac{5}{6}\right)\nn\\
+3\,\frac{\left({g_1}^2+{g_2}^2\right)^2}{16}\phi^4\left({\rm ln}
\left(\frac{\frac{1}{4}\left({g_1}^2+{g_2}^2\right)\phi^2}{\mu^2}\right)
-\frac{5}{6}\right)
-12\,g^4\phi^4\left({\rm ln}\frac{g^2\phi^2}{\mu^2}-\frac{3}{2}\right) \Bigg],
\eea 

\nin
where $g_1$ and $g_2$ are the weak interaction coupling constants, 
while $g$ is the top--Yukawa coupling.

To have a well defined comparison between our criterium and the usual 
one, we have chosen to follow the work of Casas, 
Espinosa, and Quir\'{o}s \cite{quiro1, quiro2}. 
In particular, we have taken their boundary conditions 
for $g_1$, $g_2$, $m_t$, ... at the scale $M_{_Z}$ as well as their 
matching conditions for the determination of the physical Higgs and 
top mass (see Appendix B and \cite{quiro1,quiro2} for details). 

The RG improved potential, $V_{_{RGI}}$, is obtained following the same
steps of section 3. Naturally, the appropriate beta functions to consider 
in the RG equations are now the SM ones.  As in \cite{quiro1, quiro2}, we have 
used the two--loops beta functions \cite{jones}. Note also that, differently 
from our simpler model, we now have three additional RG equations, namely  
for $g_1$, $g_2$ and $g_S$ (the strong coupling), and that no analytic 
solution for the running of the couplings can be found. 
Choosing $t=\frac{1}{2}\ln \frac{\phi^2}{\mu^2}$, we get:  

\be\label{potRGI}
V_{_{RGI}}(\phi)=m^2(t)\frac{\phi^2(t)}{2}+
\la_{eff}(t)\frac{\phi^4(t)}{24}+\Omega(t)\,,
\ee

\nin
where $\Omega(t)$ is the scale dependent vacuum energy, 
$\phi(t)=\xi(t)\,\phi$, with 
$\xi(t)=\exp \Big(-\int_0^t\gamma(t')dt'\Big)$ and  
$\gamma(t)$ being the Higgs anomalous dimension, and
$\la_{eff}(t)$ is given by:

\bea
\la_{eff}(t)=\la+\frac{3}{8\pi^2}&\Bigg[&6\,\frac{{g_1}^4}{16}\left({\rm ln} 
\left(\frac{{g_1}^2}{4}\right)-\frac{5}{6}\right)  
-12\,g^4\left({\rm ln}\,g^2-\frac{3}{2}\right)\nn\\
&+&3\,\frac{\left({g_1}^2+{g_2}^2\right)^2}{16}\left({\rm ln}
\left(\frac{{g_1}^2+{g_2}^2}{4}\right)-\frac{5}{6}\right) \Bigg]\,,
\eea

\nin
with $\la=\la(t)$, $g=g(t)$, $g_1=g_1(t)$, $g_2=g_2(t)$.

\begin{table}[Ht]
 \begin{center}
 \begin{tabular}{cccc}
  \hline\noalign{\smallskip}
   $\Lambda\,({\rm Tev})$ & $M_{_H}^{inf}\,({\rm Gev})$ & 
   $M_{_H}\,({\rm Gev})$ & $\Delta M_{_H}\,({\rm Gev})$ \\
  \noalign{\smallskip}\hline\noalign{\smallskip}
   $1      $ & $66    $ & $55.5  $ & $10.5$ \\
   $5      $ & $88    $ & $81    $ & $7   $ \\
   $10     $ & $94.5  $ & $88.5  $ & $6   $ \\
   $100    $ & $108.5 $ & $105.5 $ & $3   $ \\
   $1000   $ & $117   $ & $115   $ & $2   $ \\
   $10^{16}$ & $137.5 $ & $137.5 $ & $0   $ \\
   \noalign{\smallskip}\hline
 \end{tabular}
 \caption{Lower bounds on the Higgs mass as a function of the 
 physical cutoff. The values of the physical parameters are 
 chosen according to \cite{quiro1,quiro2} (see also Appendix B).
 The second and third columns contain the bounds
 obtained with the convexity and instability criterium respectively.}
 \end{center}
 \label{tab1}       
\end{table}

First, we have checked that, when the usual $V_{_{RGI}} = 0$ 
criterium is used, meaning that the scale of new physics, $\La$, 
is determined as the value of $\phi$ where \cite{quiro2}:

\be\label{condy}
\la_{eff}+12\frac{m^2}{{\xi}^2\La^2}+24\frac{\Omega}{{\xi}^4\La^4}=0 \, ,
\ee

\nin
the results of \cite{quiro1, quiro2} are recovered.
Then, we have derived the physical cutoff according to our 
criterium, i.e. we have looked for the location of the external inflection 
point of $V_{_{RGI}}$. 

In Table 1 we summarise the results obtained with these two  
criteria for different values of $\Lambda$. For small cutoffs, 
the lower bounds on $M_{_H}$ given by our criterium are
$\sim 10$ ${\rm Gev}$ larger than the current determinations \cite{quiro2}, 
while for increasing values of $\Lambda$ the difference tends to disappear. 

The convergence between these two methods (for large cutoffs) has 
a simple explanation. 
Let us neglect, for a moment, the convexity constraint. As $M_{_H}$
increases, the location of the inflection point moves to higher and higher
values of $\phi$. The same is, obviously, true for the point where the 
potential vanishes. In this region, $V_{_{RGI}}$ is very 
well approximated by Eq.(\ref{vrgi}) and 
$\overline{\lambda}(\phi)$ changes very slowly with $\phi$. Therefore, 
the two criteria practically give one and the same value for $\Lambda$.

The scope of Table 1 is to provide a comparison between  
the two different methods for the determination of lower bounds 
on $m_{_H}$. To this end, the 
values of the physical parameters have been chosen according to 
\cite{quiro1, quiro2} (see Appendix B) rather than to their more 
recent measured values. The reader can easily verify that the 
results we have found with the usual criterium (reported in the 
third column of Table 1) agree with those of \cite{quiro1, quiro2}. 

Now, considering the updated values: $M_Z=91.2$ Gev, $M_W=80.4$ Gev, 
$\alpha_s=0.119$ \cite{pdg}
and $M_t=178$ Gev \cite{teva}, we find for the lower bounds on  
$M_{_H}$ the results reported in Table 2. Note that, 
taking into account the present experimental uncertainty 
on $M_t$ \cite{teva},  $M_t=178 \pm 4.3$, we get: $M_{_H} =68.5^{+3}_{-3.5} $
for $\La=1$ Tev up to $M_{_H} =143.5 \pm 8.5 $ for $\La=10^{19}$ Gev.

\begin{table}[Ht]
 \begin{center}
 \begin{tabular}{cccc}
  \hline\noalign{\smallskip}
   $\Lambda\,({\rm Tev})$ & $M_{_H}^{inf}\,({\rm Gev})$ & 
   $M_{_H}\,({\rm Gev})$ & $\Delta M_{_H}\,({\rm Gev})$ \\
  \noalign{\smallskip}\hline\noalign{\smallskip}
   $1      $ & $68.5 $ & $57.5 $ & $11  $ \\
   $5      $ & $91.5 $ & $84   $ & $7.5 $ \\
   $10     $ & $98   $ & $92   $ & $6   $ \\
   $100    $ & $113  $ & $109.5$ & $3.5 $ \\
   $1000   $ & $122  $ & $120  $ & $2   $ \\
   $10^{16}$ & $143.5$ & $143.5$ & $0   $ \\
   \noalign{\smallskip}\hline
 \end{tabular}
 \caption{Lower bounds on the Higgs mass as a function of the 
 physical cutoff. Differently from Table 1, the physical 
 parameters have been chosen according to their most recent 
 experimental determinations (see text). As for Table 1, 
 the second and third columns contain the bounds
 obtained with the convexity and instability criterium respectively.}
 \end{center}
 \label{tab2}       
\end{table}

\section{Wilsonian RG}

In the previous section we have considered a phenomenological 
application of our findings. Now, to further support our results, 
we come back to the simpler Higgs--Yukawa model of Eq.(\ref{lagra})
and show that, with the help of the Wilsonian RG method, our analysis 
can be extended beyond perturbation theory.  

For the Euclidean Wilsonian action of our model at the running scale $k$, 
$S_k[\phi,{\overline\psi}, \psi]$, we consider the following non-perturbative 
ansatz \cite{love}:

\be\label{lpa}
S_k[\phi, {\overline\psi}, \psi]=
\int d^4x \left(\frac12\partial_\mu\phi\partial_\mu\phi 
+ {\overline\psi}\gamma_\mu\partial_\mu\psi + 
U_k(\phi, {\overline\psi}, \psi)\right).
\ee

As for the case of the scalar theory (see section 2 and \cite{alex}), 
for the internal region we expect that from Eq.(\ref{lpa}) a 
non-perturbative flow equation can be obtained which reproduces the 
Maxwell construction. Here, however, our scope is to investigate 
the possibility of having an instability of the scalar potential in 
the region beyond the minimum. Therefore, we only consider 
this region, where the running for the Wilsonian potential 
of our model is given by the non-perturbative RG equation \cite{love}:

\bea\label{rgtot}
\frac{\partial U_k(\phi, \sigma)}{\partial k}=-\frac{k^3}{16\pi^2}\,\ln\left(
\frac{k^2+{U_k}^{''}(\phi,\sigma)}{k^2+{U_k}^{''}(0,\sigma)}\right)+\frac{k^3}{4\pi^2}
\,\ln\left(1+\frac{{\dot{U_k}}^2(\phi,\sigma)}{k^2}\right) \nonumber \\
-\frac{k^3}{16\pi^2}\,\ln\left(1+\frac{2\,\sigma\, \dot{U_k}(\phi,\sigma)}
{k^2+{\dot{U_k}}^2(\phi,\sigma)}\left(\ddot{U_k}(\phi,\sigma)-
\frac{{\dot{U'_k}}^2(\phi,\sigma) }{k^2+{U_k}^{''}(\phi,\sigma)} \right) \right)\, .
\eea

\nin
Here  $\sigma =\overline\psi\psi$, the prime indicates the
derivative w.r.t. $\phi$ and the dot the derivative w.r.t. $\sigma$.

The bare value of the potential, which is nothing but the boundary condition for 
the RG equation (\ref{rgtot}), is (see Eq.(\ref{lagra})):

\be\label{barep}
U_{_\Lambda}(\phi, \sigma)=\frac{1}{2} m_{_\Lambda}^2 \phi^2
+\frac{\lambda_{_\Lambda}}{24}\phi^4 +g_{_\Lambda} \phi \sigma .
\ee

\nin
We now consider 
for $U_k(\phi, \sigma)$ the additional truncation: 

\be\label{ansa2}
U_k(\phi, \sigma) = V_k(\phi) + G_k(\phi) \sigma \, ,
\ee

\nin
which means that we neglect the contributions from higher 
powers of $\overline\psi\psi$. 

Inserting Eq.(\ref{ansa2}) in Eq.(\ref{rgtot}), we finally get the 
RG equations:

\bea\label{twoeq}
\frac{\partial V_k(\phi)}{\partial k}&=&-\frac{k^3}{16\pi^2}\,\ln\left(
\frac{k^2+{V_k}^{''}(\phi)}{k^2+{V_k}^{''}(0)}\right)+\frac{k^3}{4\pi^2}\,\ln\left(
1+\frac{{G_k}^2(\phi)}{k^2}\right) \nn\\
\frac{\partial G_k(\phi)}{\partial k}&=&-\frac{k^3}{16\pi^2} 
\frac{1}{k^2+{V_k}^{''}(\phi)} \left({G_k}^{''}(\phi)-\frac{2\,
G_k(\phi)\,{G_k}^{'}(\phi)}{k^2+{G_k}^2(\phi)}\right)\, .
\eea

From Eq.(\ref{barep}) is clear that the boundary conditions for $V_k$
and $G_k$ are:
\bea\label{c23}
V_{_\Lambda}(\phi)&=&\frac{1}{2} m_{_\Lambda}^2 \phi^2+\frac{1}{24} 
\lambda_{_\Lambda} \phi^4 \nn\\
G_{_\Lambda}(\phi)&=&g_{_\Lambda} \phi \, .
\eea

\begin{figure}[Ht]
\begin{center}
\includegraphics[width=7cm,height=11cm,angle=270]{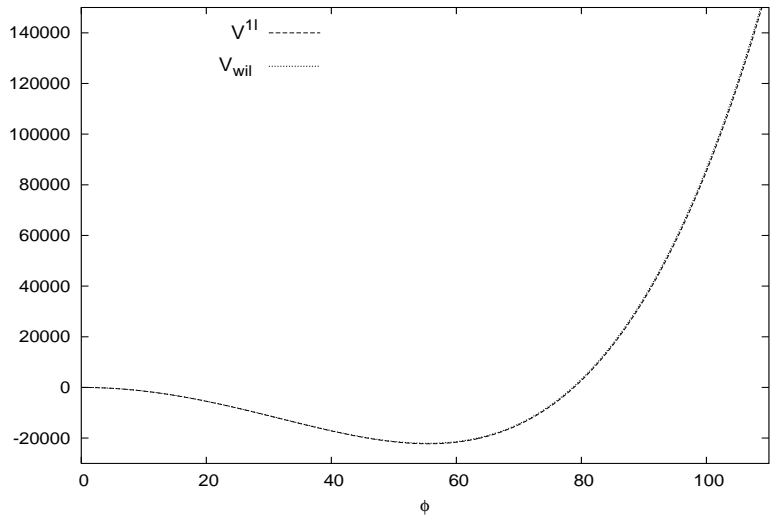}
\end{center}
\caption[]{The Wilsonian, 
$V_{wil}=V_{k=0}$, effective potential. The boundary 
values of the parameters are as in Fig.\ref{fig6}.
Only the region external to the minimum has to be considered. 
For comparison we have also plotted the one-loop 
effective potential of Fig.\ref{fig6}. We see that, 
as explained in the text, $V_{wil}$ and $V^{1l}$ are very close 
one to the other.}
\label{fig10}
\end{figure}

Given~ $m_{_\Lambda}^2$, $\lambda_{_\Lambda}$ 
and $g_{_\Lambda}$ at $k=\Lambda$, we can run the RG equations 
(\ref{twoeq}) to get for the scalar effective potential, $V_{eff}(\phi)$, 
the non-perturbative approximation: $V_{_{wil}}(\phi)=V_{k=0}(\phi)$. 
Choosing: $\lambda_{_\Lambda}=5\cdot 10^{-2}$, 
$m^2_{_\Lambda}=-1\cdot 10^{-2}$, $g_{_\Lambda}=5\cdot 10^{-1}$ at $\Lambda=100$, 
i.e. taking the same values used in fig.\ref{fig6}, we get for
$V_{_{wil}}$ the result plotted in fig.\ref{fig10} (we remind that 
the RG equation (\ref{rgtot}) is valid only in the external region).

For comparison, we have also plotted the corresponding 
$V^{1l}$ (which is nothing but the potential of 
fig.\ref{fig6}). As we can easily see, $V_{_{wil}}$ and  
$V^{1l}$ are very close one to the other. This result could have 
been guessed. 
As we have already said, in fact, in the external region the path integral 
that defines the effective potential is dominated by a single saddle 
point. As a consequence, we expect that the loop-expansion, and in particular 
the one-loop potential, provides a good approximation for $V_{eff}$.
The close coincidence between $V^{1l}$ (perturbative) and $V_{_{wil}}$ 
(non-perturbative) supports this expectation.

\begin{figure}[Ht]
\begin{center}
\subfigure[]{\includegraphics[width=7cm,height=6.5cm,angle=270]{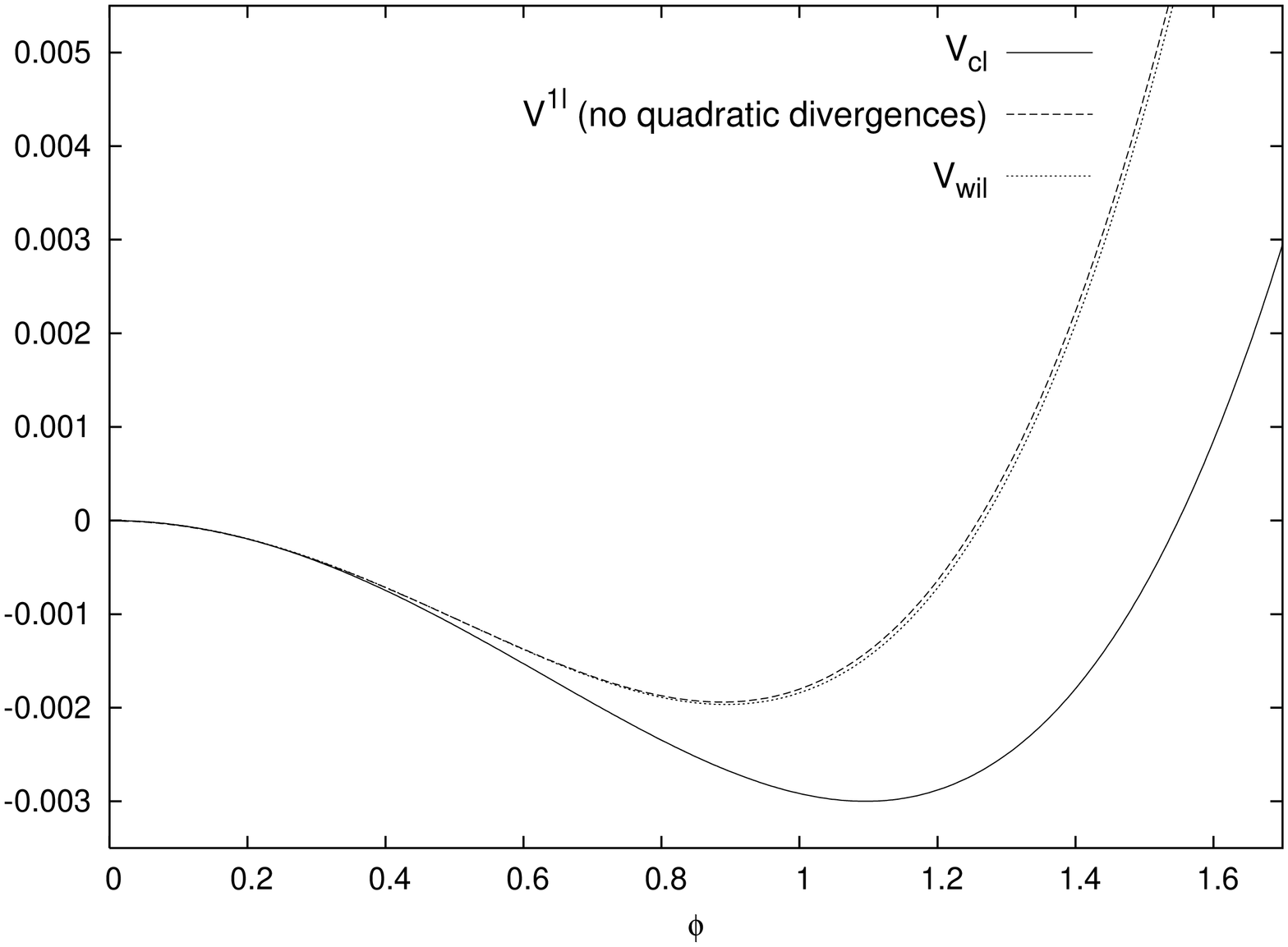}}
\subfigure[]{\includegraphics[width=7cm,height=6.5cm,angle=270]{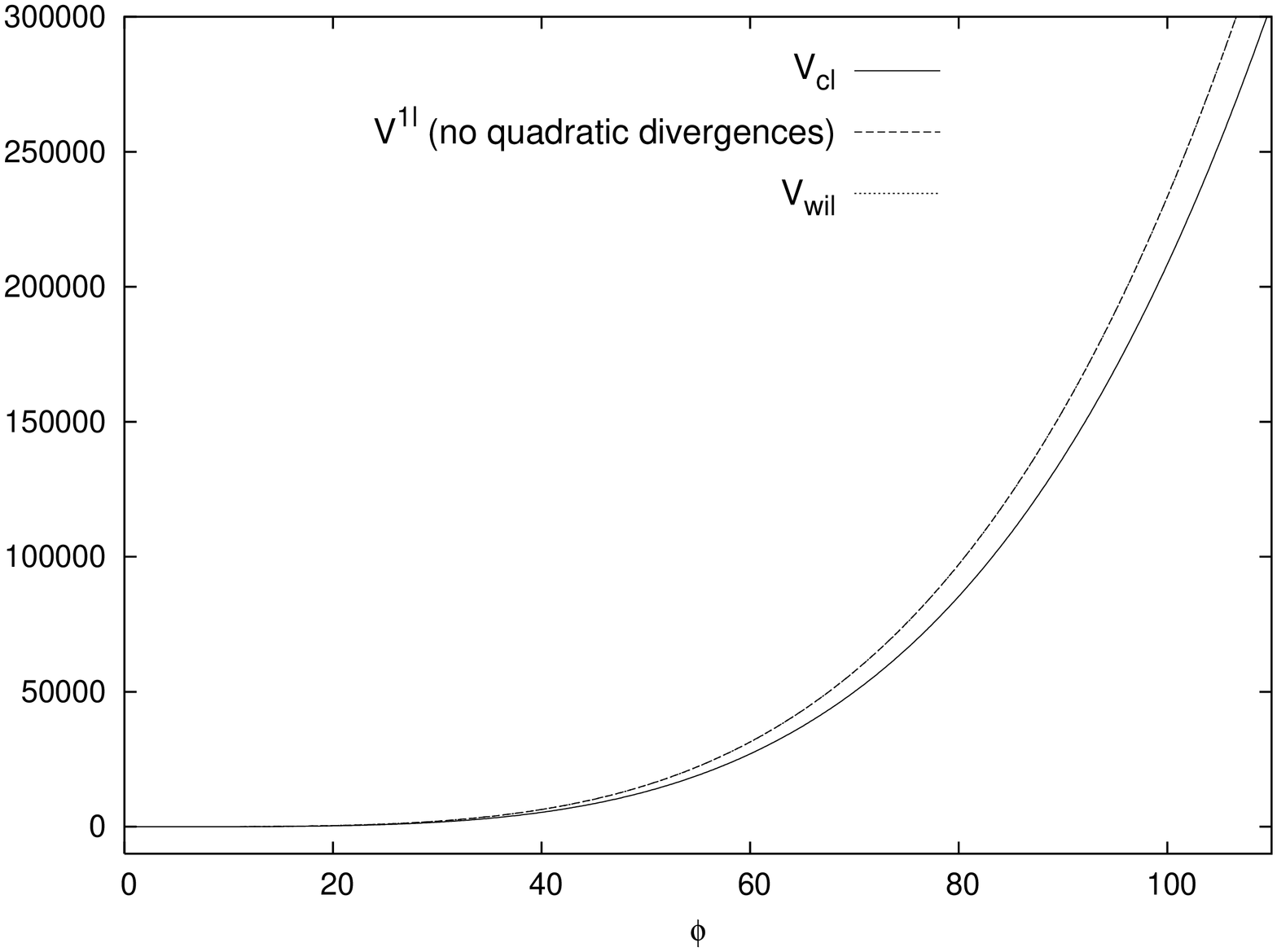}}
\end{center}
\caption[]{The bare together with the Wilsonian potential
after subtraction of the quadratic divergences. The boundary 
values of the parameters are as in Fig.\ref{fig6}. For $V_{wil}$,
only the external region has to be considered. For comparison, we have also 
plotted the one-loop effective potential of Fig.\ref{fig7}.
We see that, even after subtracting the quadratic divergences, $V_{wil}$
and $V^{1l}$ are quite close. }
\label{fig8}
\end{figure}

By its own construction, the Wilsonian method does not contain any ad hoc 
subtraction of terms. This is why we have compared the effective potential 
found with Eqs.(\ref{twoeq}) with the original one-loop result, the potential 
of fig.\ref{fig6}, where the quadratically divergent terms were kept.

If we want to make contact with the perturbative $V^{1l}$ where 
the quadratic divergences are subtracted (fig.\ref{fig7}), we 
need to implement this operation in the flow equations.

Performing a polynomial expansion of $V_k(\phi)$ and $G_k(\phi)$, we 
easily see that the subtraction of the quadratic divergences in our 
flow equations amounts to add the term: 

\be\label{count}
\left(-\frac{\lambda_k}{32\pi^2}
+\frac{{g_k}^2}{4\pi^2}\right) k\,\phi^2
\ee
 
\nin
to the first of Eqs.(\ref{twoeq}). 
In Eq.(\ref{count}), $\lambda_k$ is the coefficient of 
$\phi^4$ in the expansion of $V_k(\phi)$, while 
$g_k$ is the coefficient of $\phi$ in the expansion of $G_k(\phi)$.
Moreover, at each step of the RG iteration, $\lambda_k$ and $g_k$ are 
determined via a polynomial fit of $V_k$ and $G_k$ respectively. Their 
boundary values, of course, are $\lambda_{_\Lambda}$ and $g_{_\Lambda}$.   

Taking for $m^2_{_\Lambda}$, $\lambda_{_\Lambda}$,  $g_{_\Lambda}$ 
and $\Lambda$ the same values considered above, we now run the modified system 
of RG equations and get for $V_{_{wil}}$  the result plotted in 
figs.\ref{fig8}.
As before, we note that $V_{wil}$ and  $V^{1l}$ are very close.  

The results of the present section strongly support our 
previous findings. Even within the non-perturbative framework considered 
here, the effective potential does not show any sign of instability. 

\section{Summary and conclusions}

Starting with the analysis of some popular, but misleading,  
arguments, we have studied the instability 
problem of the EW vacuum with the help of a Higgs--Yukawa model. 

Combining the Bogolubov approach to symmetry breaking, 
namely the criterium of dynamical instability, 
with the Wilsonian RG method, we have shown that there is no  
conflict between the convexity of the effective potential (effective 
action) and the existence of broken phase vertex functions. This 
preliminary step was helpful in establishing the incorrectness of 
the above quoted arguments. 

Successively, we have shown that the potential instability is due to 
an illegal extrapolation of the renormalised effective potential into 
a region where the results of renormalised perturbation theory do not 
hold. Moreover, in agreement with what is expected from general 
theorems, we have found that the effective potential of the cutoff 
Higgs--Yukawa model is convex allover the region where is defined. 
 
To establish these results, it was necessary to go beyond the usual
application of the perturbation theory conditions. In this 
respect, we note that the dimensional regularization scheme, by its 
own construction, directly gives the results of renormalised perturbation theory. 
As the subject of this paper shows, however, the connection 
between the UV and the IR sector of the theory (the relation between 
bare and renormalised theory) can present aspects which are 
hidden to a naive application of dimensional regularization. 

In our case, the consistency constraint for the theory ($\phi \leq \Lambda$) 
and Eq.(\ref{newrg3}) imply that the combination 
$\frac{\lambda_{_{v}}}{24} + \frac{g^4}{16\pi^2}\,{\rm ln}
\frac{v^2}{\phi^2}$ cannot be negative. When we blindly jump to the 
perturbation theory results, this information  is lost. Actually, 
Eqs.(\ref{pertu}) and 
(\ref{pertur}), typically considered as the only conditions for the renormalised 
perturbation theory to hold, do not contain the above independent constraint. 
The effective potential appears to 
be unstable when this condition is ignored. 

We started our analysis within the framework of the momentum cutoff 
regularization scheme. Successively, with the help of the Wilsonian 
RG method, our results were established in a more general 
non-perturbative context. 

Moreover, despite the stability of the potential, we have shown that
lower bounds on the Higgs mass can still be derived. In fact, for a given 
renormalised value of $\lambda$, the corresponding cutoff can be found 
looking for the inflection point of $V_{eff}$ in the external region 
($\phi > v$). If the scale of new physics is not too high, a sizeable 
difference between our bounds and the usual ones is obtained. 
For $\Lambda$ in the Tev region, we find a value of $m_{_H}$ which is 
some $10-11$ Gev higher than the current determination.

In addition to these phenomenological applications, 
it is worth to note that there is a deep conceptual difference between 
our analysis and the usual one. While in our case the stability of the 
potential, as well as 
the bounds on $m_{_H}$, come as a manifestation of the 
internal consistency of the theory, in the usual approach the bounds 
are the result of an (apparently) additional constraint to be imposed 
on the potential, the requirement of stability.   
The instability is considered as a theoretically 
legitimate possibility. In fact, the meta-stability scenario explores
the consequences of having a minimum lower than the EW one. Our results 
exclude this scenario.

In the present work we have been interested on the instability 
issue only. However, we believe that our results come as a manifestation of a  
general problem, the (somehow delicate) connection between 
the UV and the IR sector of a theory and that a similar analysis can be 
applied to other cases where this connection is expected to play an 
important role. We hope to come to this point in the future.

As already said, we come now to the comparison 
of our work with \cite{kuti, holland}. 
First of all we note that the instability problem concerns the 
renormalised effective potential. Therefore, it is important to 
perform the analysis within a range of $\phi$ where renormalised 
perturbation theory is (or is supposed to be) valid. In 
\cite{kuti, holland}, however, the potential has a minimum at 
the cutoff, i.e. at $\phi \sim \Lambda = \frac{\pi}{a}$ (see fig.2 
of \cite{kuti} and fig.4 of \cite{holland}) and all the relevant 
scales, namely the ``low energy scale" $\mu$, the cutoff scale 
$\Lambda$ and the minimum $v$ are  of the same order. 
In our opinion, this hardly helps in understanding
the origin of the instability problem.

Moreover, the renormalised potential (see Eq.(2) in \cite{holland}) is obtained 
from the (subtracted) bare potential (see Eq.(5) in \cite{holland}) 
after expanding in $\frac{\phi}{\Lambda}$ and neglecting negative 
powers of $\Lambda$. Insisting on the difference between 
the bare and the renormalised potential for values of $\phi$ beyond 
$\Lambda$, as done by the authors, once more does not help in clarifying 
the problem.

We believe that we have clearly identified the origin of the 
apparent instability of the effective potential. Contrary to what 
is stated in \cite{holland}, it seems to us that it 
has nothing to do with the triviality of the theory. 

Finally, we note that in \cite{kuti, holland}, in order to avoid 
problems with the convexity of $V_{eff}$ (as stated by the authors), 
the constrained potential is used. On the contrary, insisting 
on the convexity of $V_{eff}$ as a guiding property, we have found the 
flaw that artificially makes $V_{eff}$ unstable 
in the external region. 

\vskip 20 pt
\nin
{\Large{\bf Acknowledgments}}
\vskip 10 pt

\nin
We would like to thank V. Bernard, G.Veneziano, M. Winter for many 
helpful discussions.

\newpage

\appendix{}\label{aaa}

\section {Renormalization conditions}

In this Appendix we compute the renormalised potential of 
Eq.(\ref{newrg}), where the renormalization conditions 
that keep the minimum and the curvature around the minimum fixed 
at their classical values are implemented. Clearly, these conditions are:

\bea 
\label{mini}  \left(\frac{d V^{1l}}{d \phi}\right)_{\phi=v}&=&0  \\
\label{curv}  \left(\frac{d^2 V^{1l}}{d \phi^2}\right)_{\phi=v}
&=&\frac{\la v^2}{3}=-2 m^2 \, ,
\eea

\nin
with $v=\sqrt{\frac{-6 m^2}{\la}}$. From Eq.(\ref{eff2}) we get:

\bea \label{dU}
\frac{d V^{1l}}{d \phi}&=&\phi\,\Bigg(m^2+\delta m^2+(\la+\delta\la)
\frac{\phi^2}{6}+\left(\frac{\la}{32 \pi^2}-
\frac{g^2}{4 \pi^2}\right)\La^2 \nn\\
&+&\frac{\la}{32 \pi^2}\left(m^2+\frac{\la}{2} \phi^2\right)
\ln \frac{m^2+\frac{\la}{2} \phi^2}{\La^2}-\frac{g^4 \phi^2}{4 \pi^2}\ln 
\frac{g^2 \phi^2}{\La^2} \Bigg)\, ,
\eea

\nin
so that the condition (\ref{mini}) becomes:

\bea\label{eq1}
0&=&\de m^2+\de\la\,\frac{v^2}{6}+\left(\frac{\la}{32 \pi^2}-
\frac{g^2}{4 \pi^2}\right)\La^2 \nn \\ 
&+&\frac{\la}{32 \pi^2}\left(m^2+\frac{\la}{2} v^2
\right) \ln \frac{m^2+\frac{\la}{2} v^2}{\La^2}
-\frac{g^4 v^2}{4 \pi^2}\ln \frac{g^2 v^2}{\La^2}\, .
\eea

\nin
Deriving  $V^{1l}$ once more w.r.t. $\phi$, we get:

\bea\label{d2U}
\frac{d^2 V^{1l}}{d \phi^2}&=&m^2+\delta m^2+
\left(\frac{\la}{32 \pi^2}-\frac{g^2}{4 \pi^2}
\right)\La^2+\frac{\la}{32 \pi^2}\left(m^2+\frac{3}{2}\la \phi^2\right) \ln \frac{m^2+
\frac{\la}{2}\phi^2}{\La^2}\nn\\
&+&\frac{\phi^2}{2}\left(\la+\delta \la + \frac{\la}{16 \pi^2}-
\frac{3 g^4}{2 \pi^2} \left( \ln \frac{g^2 \phi^2}{\La^2}+\frac{2}{3}\right)\right)\, , 
\eea

\nin
and the condition (\ref{curv}) reads:
\bea\label{eq2}
0&=&\de m^2+\de\la\,\frac{v^2}{2}+\left(\frac{\la}{32 \pi^2}-\frac{g^2}{4 \pi^2}
\right)\La^2+\frac{\la}{32 \pi^2}\left(m^2+\frac{3}{2}\la v^2\right) \ln \frac{m^2+
\frac{\la}{2}v^2}{\La^2} \nn\\
&+&\frac{v^2}{2}\left(\frac{\la^2}{16 \pi^2}-\frac{3 g^4}{2 \pi^2} \left( \ln
\frac{g^2 v^2}{\La^2}+\frac{2}{3}\right)\right)\, .
\eea

\nin
From Eqs.(\ref{eq1}) and (\ref{eq2}) we find:
\bea
\de \la&=&\frac{3 g^4}{2 \pi^2}\left(\ln \frac{g^2 v^2}{\La^2}+1\right)
-\frac{3\la^2}{32 \pi^2}\left(\ln \frac{m^2+\frac{\la}{2} v^2}
{\La^2}+1\right)\label{deltal}\\
\de m^2&=&\left(\frac{g^2}{4 \pi^2}-\frac{\la}{32 \pi^2}\right)\La^2
-\frac{\la m^2}{32 \pi^2}\left(
\ln \frac{m^2+\frac{\la}{2} v^2}{\La^2}+3\right)
-\frac{g^4 v^2}{4 \pi^2}\, .\label{deltam}
\eea

\nin
Inserting Eqs.(\ref{deltal}) and (\ref{deltam}) in $V^{1l}$, i.e.   
in Eq.(\ref{eff2}), we finally get:

\bea\label{Ufix}
V^{1l}&=&\frac12m^2 \phi^2 +\frac{\la}{24} \phi^4 +\frac{\left(m^2+\frac{\la}{2} 
\phi^2\right)^2}
{64 \pi^2} \left( \ln \frac{m^2+\frac{\la}{2} \phi^2}{m^2+\frac{\la}{2} v^2}
-\frac{3}{2}\right) \nn\\
&-&\frac{g^4 \phi^4}{16 \pi^2}\left(\ln \frac{\phi^2}{v^2}-\frac{3}{2}\right)
+\frac{v^2}{2}\phi^2\left(
\frac{3\la^2}{32 \pi^2}-\frac{g^4}{4 \pi^2}\right)\, .
\eea

\nin
Finally, neglecting the bosonic contribution to the quantum fluctuation 
determinant, we see that Eq.(\ref{Ufix}) is nothing but the renormalised 
one-loop potential of Eq.(\ref{newrg}).

\section{RG-improved Potential for the SM}

In the present Appendix we provide some useful relations needed for the 
computation of the RG--improved one-loop effective potential of the SM 
(section 6).  
Following \cite{quiro1}, the matching conditions for the Higgs and the top 
masses are taken as: 
\bea
M_H^2(t)&=&m_H^2(t^*) \frac{\xi^2(t^*)}{\xi^2(t)}
+Re\left(\Pi(p^2=M_H^2)-\Pi(p^2=0)\right)\label{tstar} \\
M_t&=&m_t(M_t)\left(1+\frac{{g_s(M_t)}^2}{3\pi^2}\right) \,, 
\eea
where $\Pi$ is the self--energy of the Higgs boson (for the full explicit 
expression see the appendix A of \cite{quiro1}). Moreover, although the 
exact effective potential is scale independent, for $V^{1l}$ and 
$V_{_{RGI}}$ this is true only approximately.
The value $t^*$ of the parameter $t$ that appears in Eq.(\ref{tstar})
is chosen as to minimize the dependence of $V_{_{RGI}}$ on the choice
of the running scale $\mu(t) = M_{_Z}e^t$. The corresponding $\mu(t^*)$, in our case, 
is:  $\mu(t^*) \sim 130$ Gev. 

Accordingly, omitting the Higgs and the Goldstone (negligible) contributions,
the value of $m_H^2(t^*)$ is secured as \cite{quiro1}:
\bea
m_H^2(t^*)&=&\xi^2(t^*)v^2\Bigg( \frac{\la(t^*)}{3}+\frac{3}{64\pi^2}\bigg\{ 
g_1^4(t^*)\bigg[\log \frac{g_1^2(t^*)\xi^2(t^*)v^2}{4\,\mu^2(t^*)}+
\frac{2}{3}\bigg] \nn \\
&+&\frac{1}{2}{\left[g_1^2(t^*)+g_2^2(t^*)\right]}^2\bigg[\log \frac{\left[g_1^2(t^*)+
g_2^2(t^*)\right]\xi^2(t^*)v^2}{4\,\mu^2(t^*)}+\frac{2}{3}\bigg] \nn\\
&-&8\,g^4(t^*)\log \frac{g^2(t^*)\xi^2(t^*)v^2}{2\,\mu^2(t^*)} \bigg\}\, \Bigg)\,, 
\eea

\nin
where in the first term of the r.h.s. we recognise the tree--level relation
for $m_H^2$, while the other terms come from the loop corrections. 

The boundary values for the coupling constants are choosen as \cite{quiro1}:
\bea
g_1(M_Z)&=&0.650 \nn\\
g_2(M_Z)&=&0.355 \nn\\
g_s(M_Z)&=&1.218 \nn\\
\gamma(M_Z)&=&0  \nn\\
\Omega(M_Z)&=&0  \nn\\
g(M_t)&=&\frac{\sqrt{2}\,m_t(M_t)}{\xi(M_t)\,v}=0.9635 \, ,
\eea

\nin
which correspond to $M_W=80$ Gev, $M_Z=91.2$ Gev, $\alpha_s=0.118$ and
$M_t=175$ Gev. 

The coupling $\lambda(M_Z)$ is kept as a free parameter. 
As explained in the text, by considering different values of 
$\lambda(M_Z)$, we obtain different values for the physical cutoff,  
thus getting lower bounds for the Higgs mass as a function of the scale 
of new physics.

Note also that, in order to keep the location of the minimum to its 
phenomenological value, $m^2$ has to be fixed by the condition: $\frac{
<\phi(t^*)>}{\xi(t^*)}=v=246.22$ Gev, which gives \cite{quiro1}: 

\bea
m^2(t^*)&=&-\xi^2(t^*)v^2\Bigg( \frac{\la(t^*)}{6}+\frac{3}{64\pi^2}\bigg\{ 
\frac{1}{2}g_1^4(t^*)\bigg[\log \frac{g_1^2(t^*)\xi^2(t^*)v^2}{4\,
\mu^2(t^*)}-\frac{1}{3}\bigg] \nn \\
&+&\frac{1}{4}{\left[g_1^2(t^*)+g_2^2(t^*)\right]}^2\bigg[\log \frac{\left[g_1^2(t^*)+
g_2^2(t^*)\right]\xi^2(t^*)v^2}{4\,\mu^2(t^*)}-\frac{1}{3}\bigg] \nn\\
&-&4\,g^4(t^*)\bigg[\log \frac{g^2(t^*)\xi^2(t^*)v^2}{2\,
\mu^2(t^*)}-1\bigg] \bigg\}\, \Bigg) \, . 
\eea

Now, solving numerically the system of RG equations for the running coupling 
constants, we get Eq. (\ref{potRGI}) of section 6 for $V_{_{RGI}}(\phi)$. 

We end this appendix giving the boundary values of the
coupling constants corresponding to the updated values of $M_Z$, $M_W$,
$\alpha_{_S}$ and $M_{_t}$ reported in section 6: 
$g_1(M_Z)=0.653$, $g_2(M_Z)=0.349$, $g_s(M_Z)=1.223$ and $g(M_t)=0.980$.   

\end{document}